\documentclass[twocolumn, prc, amssymb, superscriptaddress, aps,
preprintnumbers,amsmath,floatfix]{revtex4}
\usepackage{multirow}
\usepackage{isotope}
\usepackage{supertabular}
\usepackage{graphicx,amsfonts,color,epsfig}
\usepackage{amsmath}
\usepackage{amssymb}
\usepackage{url}

\begin{document}


\title{$^{229m}$Th isomer from a nuclear model perspective}


\author{Nikolay \surname{Minkov}}
\email{nminkov@inrne.bas.bg} \affiliation{Institute of Nuclear Research and
Nuclear Energy, Bulgarian Academy of Sciences, Tzarigrad Road 72, BG-1784
Sofia, Bulgaria} \affiliation{Max-Planck-Institut f\"ur Kernphysik,
Saupfercheckweg 1, D-69117 Heidelberg, Germany}

\author{Adriana P\'alffy}
\email{Adriana.Palffy-Buss@fau.de}
\affiliation{Max-Planck-Institut f\"ur Kernphysik, Saupfercheckweg 1,
D-69117 Heidelberg, Germany}
\affiliation{Department of Physics, Friedrich-Alexander-Universit\"at
Erlangen-N\"urnberg,  D-91058 Erlangen, Germany}


\date{\today}

\begin{abstract}
The physical conditions for the emergence of the extremely low-lying
nuclear isomer $^{229m}$Th at approximately 8~eV are investigated in the
framework of our recently proposed nuclear structure model. Our theoretical
approach explains the $^{229m}$Th-isomer phenomenon as the result of a very
fine interplay between collective quadrupole-octupole and single-particle
dynamics in the nucleus. We find that the isomeric state can only appear in
a rather limited model space of quadrupole-octupole deformations in the
single-particle potential, with the octupole deformation being of a crucial
importance for its formation. Within this deformation space the
model-described quantities exhibit a rather smooth behaviour close to the
line of isomer-ground state quasi-degeneracy determined by the crossing of
the corresponding single-particle orbitals. Our comprehensive analysis
confirms the previous model predictions for reduced transition
probabilities and the isomer magnetic moment, while showing a possibility
for limited variation in the ground-state magnetic moment theoretical
value. These findings prove the reliability of the model and suggest that
the same dynamical mechanism could manifest in other actinide nuclei giving
a general prescription for the search and exploration of similar isomer
phenomena.
\end{abstract}

\maketitle

\section{Introduction}
\label{intro}

Well supporting the current strong emphasis on  interdisciplinary research, a
unique extremely low-lying $^{229m}$Th isomer at approximately~8 eV
\cite{Beck_78eV_2007,Beck_78eV_2007_corrected,Seiferle_EnTh229m_2019,
Yamaguchi_EnTh229m_2019} obviously disregards the recognized low-energy
border of nuclear physics firmly stepping on atomic physics territory.
Although another low-lying nuclear excitation in $^{235}$U also approaches
this limit with an order of magnitude larger energy of 76 eV
\cite{Browne_NDS_A235_2014}, currently $^{229m}$Th attracts much more
interest since its energy lies in the range of accessibility  of present
vacuum ultraviolet (VUV) lasers capable to handle the wavelength of 150 nm
($\approx 8$ eV). With a relatively narrow width and excellent stability,
this transition appears to be of a practical interest for a diverse community
beyond nuclear physics, involving atomic, laser, plasma physics, metrology,
cosmology and others, posing a number of puzzling problems and raising hopes
for possible advanced applications. The main interest is related to a new
frequency standard based on laser access and stabilization of this transition
with sufficient accuracy through contemporary laser (frequency comb or other)
techniques. This has been often referred to in the literature as a ``nuclear
clock'' \cite{Peik_Clock_2003,Peik_Clock_2015,Campbell_Clock_2012}. Such a
nuclear clock is expected to have a  better or at least comparable accuracy
to the currently developed atomic clocks. This entails a rich variety of
possible $^{229m}$Th-based applications such as the precise determination of
temporal variations in fundamental constants
\cite{Flambaum06,Berengut2009,Rellegert2010,Flambaum2020}, the development of
nuclear lasers in the VUV range \cite{Tkalya_NuclLaser_2011}, detection
improvements in satellite and deep space navigation, gravitation waves,
geodesy, precise analysis of  chemical environment and others.

Towards the aforementioned applications, recent experiments have confirmed
the existence of the isomer \cite{Wense_Nature_2016} and have determined the
isomer mean half-life in neutral Th atoms \cite{Seiferle_PRL_2017}.
Furthermore, the magnetic dipole moment $\mu$ of the nuclear isomeric state
(IS) was determined for the first time through laser spectroscopy experiments
\cite{Thielking2018,Mueller18} providing the value of $\mu_{\mbox{\scriptsize
IS}}=-0.37(6)\mu_N$. Then, three very recent experiments proposed newly
updated values for the isomer energy, $E_{IS}=8.28(17)$ eV
\cite{Seiferle_EnTh229m_2019} (from internal conversion electron
spectroscopy), $E_{IS}=8.30(92)$ eV \cite{Yamaguchi_EnTh229m_2019} (by
determining the transition rates and energies from the above level at 29.2
keV) and $E_{IS} = 8.10(17)$ eV \cite{Sikorsky2020} (from a
micro-calorimetric determination of absolute $\gamma$-ray energy
differences).

These advances, although not yet reaching the accuracy needed for a nuclear
clock, pose new challenges and inspire new studies of the $^{229m}$Th problem
from the nuclear structure side. $^{229}$Th belongs to the light actinide
nuclear mass region known for the presence of enhanced collectivity and shape
dynamic properties suggesting a complicated interaction between the
collective motion of the even-even core and the individual motion of the
single neutron. The single-particle (s.p.) states of the latter determine the
$^{229}$Th ground state (GS) with $K^{\pi}=5/2^{+}$ and the IS with
$K^{\pi}=3/2^{+}$ based on the 5/2[633] and 3/2[631] s.p. orbitals. Here,
$\pi$ denotes the parity and $K$ refers to the projection of the total
nuclear angular momentum on the body-fixed principal symmetry axis of the
system, respectively. We use the usual Nilsson notation $K[Nn_z\Lambda]$ with
$N$, $n_z$ and $\Lambda$ being the asymptotic Nilsson quantum numbers
\cite{NR1995}. Although it is intuitively clear that the entire nuclear
structure dynamics should essentially influence the appearance and the
properties of the isomer, only limited work has addressed this aspect in the
past. Thus, predictions for the $B(M1)$ and $B(E2)$ reduced
$3/2^{+}\rightarrow 5/2^{+}$ transition probabilities have been made in
Refs.~\cite{Gulda02,Ruch06} using the quasiparticle-plus-phonon model
\cite{Sol76} without particular focus on the isomer properties. Furthermore,
in Refs.~\cite{Dyk98,Tkalya15} estimates were made for the isomer $B(M1)$
transition rate using the Alaga branching ratios \cite{Alaga55}, and for the
IS magnetic moment $\mu_{\mbox{\scriptsize IS}}$ based on the Nilsson model
\cite{Nilsson1955}. The obtained value $\mu_{\mbox{\scriptsize IS}}$= -0.076
$\mu_N$ essentially differs from the recently available experimental value of
$-0.37(6)\mu_N$ \cite{Thielking2018}.

Understanding the physical mechanism behind the $^{229m}$Th phenomenon
requires a thorough investigation of the interplay of all involved collective
and s.p. degrees of freedom, and identification of all structure effects
which could allow the appearance of an excitation in the eV energy scale.
Since the latter is beyond reach for the accuracy of nuclear models generally
speaking, the implementation of such a task would require the application of
a sophisticated theoretical method which can provide the necessary conclusion
by juxtaposing results and information gained from different perspectives and
observables such as energies, transition probabilities and magnetic moments.
Motivated by the considerations above we have recently put forward a complete
nuclear-structure model approach that takes into account the axial
quadrupole-octupole (QO) (pear-shape) deformation modes typical for the
nuclei in the actinide region  both in the collective and s.p. degrees of
freedom of the nucleus \cite{Minkov_Palffy_PRL_2017}. The formalism involves
in the even-even nuclear core the so-called coherent QO model, describing
collective axial quadrupole and octupole vibrations with equal oscillation
frequencies non-adiabatically coupled to the rotation motion
\cite{b2b3mod,b2b3odd,MDSSL12,MDDSLS13}, while the odd-nucleon motion is
described within a deformed shell model (DSM) including reflection-asymmetric
Woods-Saxon potential \cite{qocsmod} and pairing correlations of
Bardeen-Cooper-Schrieffer (BCS) type with blocking of the unpaired nucleon
orbital \cite{Ring1980}. The odd-nucleon motion is coupled to the collective
motion by a Coriolis interaction taken into account through  perturbation
theory. The model spectrum has the form of quasi-parity-doublet bands built
on the ground and excited quasi-particle (q.p.) states. In this scheme the IS
appears as a q.p. band head of an excited quasi-parity-doublet. The model
framework allows a rather complete and intrinsically consistent spectroscopic
treatment of the nucleus including its IS.

Based on this model description we were able in
Ref.~\cite{Minkov_Palffy_PRL_2017} to predict the $B(M1)$ and $B(E2)$ reduced
probabilities for the IS $3/2^{+}\rightarrow 5/2^{+}$ transition. For $B(M1)$
we have provided the limits of $0.006-0.008$ Weisskopf units (W.u.), well
below the earlier deduced values of 0.048 W.u. \cite{Dyk98,Tkalya15} and
0.014 W.u. \cite{Ruch06}, corroborating the experimental difficulties to
observe radiative isomer decay
\cite{Jeet_PRL_2015,Yamaguchi2015,LarsThesis2016}. For the electric
quadrupole $B(E2)$  transition probability we have determined the limits of
$B(E2)$=20--30 W.u. At the same time the energy spectrum and several
available data on other transition rates were described with reasonable
accuracy. In the subsequent work \cite{Minkov_Palffy_PRL_2019} we have
calculated the magnetic moment of the IS, $\mu_{\mbox{\scriptsize IS}}$ and
of the GS, $\mu_{\mbox{\scriptsize GS}}$, by taking into account attenuation
effects in the spin and collective gyromagnetic factors, without changing the
model parameters originally adjusted in Ref.~\cite{Minkov_Palffy_PRL_2017}.
The result for $\mu_{\mbox{\scriptsize IS}}$ in the range from
$\mu_{\mbox{\scriptsize IS}}=-0.25\mu_N$ to $-0.35\mu_N$ is in rather good
agreement with the recent experimental values $(-0.3)$--$(-0.4)$
\cite{Mueller18} and $-0.37(6)$ \cite{Thielking2018}. On the other hand,
$\mu_{\mbox{\scriptsize GS}}$ was obtained in the range
$\mu_{\mbox{\scriptsize GS}}=0.53-0.66$ $\mu_N$, overestimating the latest
reported and older experimental values of $0.360(7)$ $\mu_N$
\cite{Safronova13} and $\mu_{\mbox{\scriptsize GS}}=0.45\mu_N$
\cite{Gerstenkorn74}, respectively, and being in agreement with an earlier
theoretical prediction $\mu_{\mbox{\scriptsize GS}}=0.54\mu_N$ based on the
modified Woods-Saxon potential \cite{Chasman1977}. Our model analysis in
Ref.~\cite{Minkov_Palffy_PRL_2019} showed that the Coriolis $K$-mixing
interaction lowers $\mu_{\mbox{\scriptsize GS}}$ pushing it towards the
experimental values, while its effect on $\mu_{\mbox{\scriptsize IS}}$ is
negligible due to the circumstance that the $K^{\pi}=3/2^{+}$ IS has no
mixing partner with angular momentum $I^{\pi}=3/2^{+}$ in the GS band.

These results raise several important questions to our understanding of the
$^{229m}$Th problem from the nuclear structure perspective, which we address
in this work. (i) To which  extent does the shape dynamics play a role for
the emergence of such a nuclear structure phenomenon as the tiny energy
difference between the IS and the GS? (ii) What is the degree of
arbitrariness in the choice of parameters providing the model predictions?
The basic input in DSM are the quadrupole ($\beta_{2}$) and octupole
($\beta_{3}$) deformations, which determine the s.p. orbitals on which the GS
and IS are formed. It is, therefore, important to identify the regions in the
($\beta_2,\beta_3$) deformation space of DSM which provide a relevant model
treatment of the isomer and the overall spectroscopic properties of the
nucleus. To clarify this question, in this work we perform DSM calculations
on a grid in a wide range in the QO deformation space covering the regions of
physical relevance for a nucleus in the actinide mass region around
$^{229}$Th. A next question that we address is (iii) whether by including the
experimental GS and IS magnetic moment values in the model fits made for
different pairs of DSM QO deformations, a better reproduction of
$\mu_{\mbox{\scriptsize GS}}$ could be achieved? How would the model
predictions for the other spectroscopic quantities and in particular for
$B(M1)$ and $B(E2)$  change?  Finally, (iv) is $^{229m}$Th  a unique
phenomenon appearing by chance, or the considered dynamical mechanism could
provide the presence of similar not yet observed phenomena in other nuclei?
In this work we aim to provide answers to these questions, prove the degree
of reliability of the model predictions, and clarify details of the mechanism
which governs the appearance of the IS.

This  work is structured as follows. Sec. \ref{dasmodel} reviews the model
formalism in a self-contained form together with details on its application
to the $^{229m}$Th problem. In Sec. \ref{results} results from the
calculations in the QO deformation space of DSM with the corresponding
behaviour of the IS energy, $B(M1)$, $B(E2)$ transition rates and the IS and
GS magnetic moments are presented and discussed. In Sec. \ref{summ} we
summarize our analysis and conclude on the reliability of the suggested model
mechanism. We thereby  provide  our updated  theoretical predictions for all
discussed observables and answer the questions formulated above.

\section{Quadrupole-octupole core plus particle model}
\label{dasmodel}

\subsection{Hamiltonian}
\label{ham}

The model Hamiltonian of axial QO vibrations and rotations coupled to the
s.p. motion with Coriolis interaction and pairing correlations can be written
in the form \cite{Minkov_Palffy_PRL_2017}
\begin{eqnarray}
H=H_{\mbox{\scriptsize s.p.}}+H_{\mbox{\scriptsize pair}}+
H_{\mbox{\scriptsize qo}}+H_{\mbox{\scriptsize Coriol}}\, .
\label{Htotal}
\end{eqnarray}
Here $H_{\mbox{\scriptsize s.p.}}$ is the single-particle (s.p.) DSM
Hamiltonian with the Woods-Saxon potential for fixed axial quadrupole,
octupole and higher multipolarity deformations $(\beta_2 , \beta_3 , \beta_4
,\beta_5 , \beta_6)$ \cite{qocsmod} providing the s.p. energies
$E^{K}_{\mbox{\scriptsize sp}}$ with given value of the projection $K$ of the
total and s.p. angular momentum operators $\hat{I}$ and $\hat{j}$,
respectively on the intrinsic symmetry axis. $H_{\mbox{\scriptsize pair}}$ is
the standard BCS pairing Hamiltonian \cite{Ring1980} which together with
$H_{\mbox{\scriptsize s.p.}}$ determines the quasi-particle (q.p.) spectrum
$\epsilon^{K}_{\mbox{\scriptsize qp}}= \sqrt{(E^{K}_{\mbox{\scriptsize
sp}}-\lambda )^{2}+\Delta^{2}}$, with the chemical potential $\lambda$ and
the pairing gap $\Delta$ determined as shown in Ref.~\cite{WM10}.
Furthermore, $H_{\mbox{\scriptsize qo}}$ describes the oscillations of the
even--even core with respect to the quadrupole ($\tilde{\beta_2}$) and
octupole ($\tilde{\beta_3}$) axial deformation variables mixed through a
centrifugal (rotation-vibration) interaction \cite{b2b3mod,b2b3odd}. Its
spectrum is obtained in an analytical form by assuming equal frequencies for
the quadrupole and octupole oscillations. The latter are known as the
coherent QO mode (CQOM) and will be discussed in more detail in Sec. IIB.
Hereafter, we distinguish the CQOM collective (dynamical) variables
$\tilde{\beta_2}$ and $\tilde{\beta_3}$ from the fixed DSM deformation
parameters $\beta_2$ and $\beta_3$ considered in this work (see Sec. IIB for
clarification).

Returning to the total Hamiltonian in Eq.~(\ref{Htotal}),
$H_{\mbox{\scriptsize Coriol}}$ involves the Coriolis interaction between the
even-even core and the unpaired nucleon \cite{b2b3odd}. It is treated as a
perturbation with respect to the remaining part of Hamiltonian (\ref{Htotal})
and then incorporated into the collective QO potential of
$H_{\mbox{\scriptsize qo}}$ defined for a given angular momentum $I$, parity
$\pi$  and s.p. bandhead projection $K_{b}$ leading to a joint effective term
\cite{NM13}
\begin{eqnarray}
H_{\mbox{\scriptsize qo}}^{IK_{b}}
&=& -\frac{\hbar^2}{2B_2}\frac{\partial^2}{\partial\tilde{\beta_2}^2}
-\frac{\hbar^2}{2B_3}\frac{\partial^2}{\partial\tilde{\beta_3}^2}+
\frac{1}{2}C_2{\tilde{\beta_2}}^{2}+
\frac{1}{2}C_3{\tilde{\beta_3}}^{2} \nonumber \\
&+& \frac{\widetilde{X}(I^{\pi},K_{b})}
{d_2\tilde{\beta_2}^2+d_3\tilde{\beta_3}^2}\, . \label{HqoK}
\end{eqnarray}
Here, $B_2$ $(B_3)$, $C_2$ $(C_3)$ and $d_2$ ($d_3$) are quadrupole
(octupole) mass, stiffness and inertia parameters, respectively. The function
$\widetilde{X}(I^{\pi},K_{b})$ determines the centrifugal term in which the
Coriolis mixing is taken into account and has the form:
\begin{eqnarray}
&&\widetilde{X}(I^{\pi},K_{b})=\frac{1}{2}\Biggl[d_0+I(I+1)-K_{b}^2
\biggr.\nonumber\\
&&+(-1)^{I+\frac{1}{2}}\left(I+\frac{1}{2}\right)
a^{(\pi ,\pi^{b})}_{\frac{1}{2}} \delta_{K_{b},\frac{1}{2}}
\nonumber\\
&&\biggl.-A \mkern-18mu\!\!\sum_{\substack {\nu \neq b\\
(K_{\nu}=\frac{1}{2}, K_{b}\pm 1)}}\mkern-18mu\!\!
\frac{\left[\widetilde{a}_{K_{\nu}K_{b}}^{(\pi ,\pi^{b})}(I)\right]^{2}}
{\epsilon^{K_{\nu}}_{\mbox{\scriptsize qp}}
-\epsilon^{K_{b}}_{\mbox{\scriptsize qp}}} \biggr]\, ,
\label{Xmix}
\end{eqnarray}
where $d_0$ determines the collective QO potential origin, $A$ is the
Coriolis mixing strength defined in Ref.~\cite{NM13} and the sum is performed
over q.p. states with energies $\epsilon^{K_{\nu}}_{\mbox{\scriptsize qp}}$
above the Fermi level. For the sum we consider in our numerical calculations
ten mixing orbitals. The quantity
$a^{(\pi,\pi^{b})}_{1/2}=\pi\pi^{b}a_{\frac{1}{2}-\frac{1}{2}}^{(\pi^{b})}$
represents the decoupling factor for the case $K_{b}=1/2$, while the
quantities $\widetilde{a}_{K_{\nu}K_{b}}^{(\pi,\pi^{b})}(I)$ represent the
Coriolis mixing factors given by
\begin{equation}
\widetilde{a}_{K_{\nu}K_{b}}^{(\pi,\pi^{b})} = \left\{   \mkern-18mu
\begin{array}{c}\sqrt{(I- K_{b}) (I+ K_{b}+1)}
a_{K_{\nu}K_{b}}^{(\pi^{b})} , \, K_{\nu}=K_{b}+1 \\
\sqrt{(I+ K_{b}) (I- K_{b}+1)}a_{K_{b}K_{\nu}}^{(\pi^{b})} ,\, K_{\nu}=K_{b}-1 \\
\ \ \ \pi\pi^{b}(-1)^{(I+\frac{1}{2})}(I+\frac{1}{2})
a_{\frac{1}{2}-\frac{1}{2}}^{(\pi^{b})} , \, K_{\nu}=K_{b}=\frac{1}{2}, \\
\end{array}\right.
\label{amixtilde}
\end{equation}
with
\begin{eqnarray}
a_{K_{\nu}K_{b}}^{(\pi^{b})}&=&\frac{P^{b}_{K_{\nu}K_{b}}}
{N_{K_{\nu}}^{(\pi^{b})}N_{K_{b}}^{(\pi^{b})}}
\langle\mathcal{F}_{K_{\nu}}^{(\pi^{b})}|\hat{j}_{+}|
\mathcal{F}_{{K_{b}}}^{(\pi^{b})}\rangle \nonumber \\
&=& \frac{P^{b}_{K_{b}K_{\nu}}}
{N_{K_{b}}^{(\pi^{b})}N_{K_{\nu}}^{(\pi^{b})}}
\langle\mathcal{F}_{K_{b}}^{(\pi^{b})}|\hat{j}_{-}|
\mathcal{F}_{{K_{\nu}}}^{(\pi^{b})}\rangle .
\label{amix}
\end{eqnarray}
The latter involve matrix elements of the s.p. operators
$\hat{j}_{\pm}=\hat{j}_x\pm i\hat{j}_y$ between the parity-projected
components of the s.p. wave functions $\mathcal{F}_{K_{b}}^{(\pi^{b})}$ of
the bandhead state and the admixing state
$\mathcal{F}_{{K_{\nu}}}^{(\pi^{b})}$. Each s.p. wave function is obtained in
DSM \cite{qocsmod} as an expansion in the axially-deformed
harmonic-oscillator basis $|Nn_z\Lambda \Sigma\rangle$ (with $\Lambda+\Sigma
=K$)
\begin{eqnarray}
\mathcal{F}_{K}=\sum_{Nn_z\Lambda}C^{K}_{Nn_z\Lambda}
|Nn_z\Lambda \Sigma\rangle . \label{ahodecomp}
\end{eqnarray}
In the case of reflection asymmetry ($\beta_3\neq0$) the wave function has a
mixed parity and can be decomposed as
$\mathcal{F}_{K}=\sum_{\pi_{\mbox{\scriptsize sp}}=\pm
1}\mathcal{F}_{K}^{(\pi_{\mbox{\scriptsize sp}})}=\mathcal{F}_{K}^{(+)}+
\mathcal{F}_{K}^{(-)}$, with the s.p. parity given by $\pi_{\mbox{\scriptsize
sp}}=(-1)^N=\pm 1$. The action of the s.p. parity operator
$\hat{\pi}_{\mbox{\scriptsize sp}}$ gives $\hat{\pi}_{\mbox{\scriptsize
sp}}\mathcal{F}_{K}= \mathcal{F}_{K}^{(+)}-\mathcal{F}_{K}^{(-)}$, and for
the parity-projected parts one has $\hat{\pi}_{\mbox{\scriptsize
sp}}\mathcal{F}_{K}^{(\pm )} =\pm \mathcal{F}_{K}^{(\pm )}$. In our approach
the projection is made with respect to the experimentally assigned good
parity $\pi^{b}$ of the bandhead s.p. state (see below). It is clear that in
the presence of octupole deformation each s.p. orbital is characterized by an
average (expectation) value of the parity determined as \cite{MDSS10}
\begin{eqnarray}
\langle \hat{\pi}_{\mbox{\scriptsize sp}}\rangle  =
\sum_{Nn_z\Lambda} (-1)^{N}|C^{K}_{Nn_z\Lambda}|^{2},
\label{avepar}
\end{eqnarray}
with the expansion coefficients calculated in the DSM. The quantity $\langle
\hat{\pi}_{\mbox{\scriptsize sp}}\rangle$ takes values in the interval
$-1\leq\langle \hat{\pi}_{\mbox{\scriptsize sp}}\rangle \leq +1$ in
dependence on the octupole $\beta_3$ and quadrupole $\beta_2$ deformations
entering the DSM.

The quantity $N_{K}^{(\pi^{b})}=\left[\left\langle\mathcal{F}_K^{(\pi^{b})}
\big|\mathcal{F}_K^{(\pi^{b})}\right\rangle \right]^{\frac{1}{2}}$ in
Eq.~(\ref{amix}) is a parity-projected normalization factor, whereas
$P_{K_{\nu'}K_{\nu}}^{b}=U_{K_{\nu'}}^{b}U_{K_{\nu}}^{b}
+V_{K_{\nu'}}^{b}V_{K_{\nu}}^{b}$ involves the BCS occupation factors. The
index $b$ corresponds to the blocked s.p. orbital on which the collective
spectrum is built. Since the BCS procedure is performed separately for each
(blocked) bandhead orbital,  the overlap integrals and the matrix elements
between states built on different bandhead orbitals involve the average of
both separate occupation factors $P_{K_{\nu'}K_{\nu}}^{bb'}=
\frac{1}{2}\left(P_{K_{\nu'}K_{\nu}}^{b}+P_{K_{\nu}K_{\nu'}}^{b'} \right)$.
The  occupation factors $U$ and $V$ and the q.p. energies
$\epsilon^{K}_{\mbox{\scriptsize qp}}$ are obtained by solving the BCS gap
equation as done in Ref.~\cite{WM10} with the pairing constant
$G=G_{\mbox{\scriptsize N/P}}$ for neutron/proton subsystems of $N$ protons
and $Z$ neutrons determined as \cite{NR1995}
\begin{eqnarray}
G_{\mbox{\scriptsize N/P}}=\frac{1}{N+Z}\left (g_{0}
\mp g_{1}\frac{N-Z}{N+Z} \right ) \, .
\label{Gnp}
\end{eqnarray}
Here the pairing parameter $g_{0}$ is considered to vary between
the values $g_{0}=17.8$ MeV, used in Ref.~\cite{WM10} for DSM plus
BCS calculations in the actinide region, and $g_{0}=19.2$ MeV,
suggested in Ref.~\cite{NR1995} for rare-earth nuclei, while
$g_{1}$ is considered to vary around the value $g_{1}=7.4$ MeV used
in both references cited above.

\subsection{Model solution, spectrum and wave functions}
\label{wf}
The spectrum which corresponds to the Hamiltonian (\ref{Htotal}) represents
QO vibrations and rotations built on a q.p. state with $K=K_{b}$ and parity
$\pi^{b}$.  It is obtained in two steps. First, the s.p. and q.p. energy
levels and wave functions are obtained through a DSM plus BCS calculation
performed for fixed $\beta_2$- and $\beta_3$- parameter values of the s.p.
Woods-Saxon potential, providing the odd-nucleon energy contribution to the
bandhead, $\epsilon^{K_{b}}_{\mbox{\scriptsize qp}}$, and the Coriolis mixing
factors $\widetilde{a}$, Eq.~(\ref{amixtilde}), in the centrifugal part
$\widetilde{X}$, Eq.~(\ref{Xmix}), of the QO Hamiltonian (\ref{HqoK}). In the
second step the collective QO vibration-rotation energies and wave functions
are obtained through the solution of the Schr\"odinger equation for the
two-dimensional potential in the collective $\tilde{\beta_2}$- and
$\tilde{\beta_3}$- variables of Hamiltonian (\ref{HqoK}). In the following we
will address in more detail this second step.

In the general case of arbitrary values of the Hamiltonian parameters $B_2$,
$B_3$, $C_2$, $C_3$ and $d_2$, $d_3$, the solution of the Schr\"odinger
equation in $\tilde{\beta_2}$  and $\tilde{\beta_3}$ has to be obtained
numerically.  A transformation of variables introduces the ellipsoidal
``radial'' and ``angular''coordinates, respectively,
\begin{eqnarray} \eta=\left[\frac{2(d_2\tilde{\beta_2}^2+d_3\tilde{\beta_3}^2)}
{d_2+d_3}\right]^{\frac{1}{2}}, \ \
\phi=\arctan\left( {\frac{\tilde{\beta_3}}{\tilde{\beta_2}}
\sqrt{\frac{d_3}{d_2}}}\right ),
\end{eqnarray}
such that
\begin{eqnarray}
\tilde{\beta_{2}}=p\eta\cos\phi, \qquad \tilde{\beta_{3}}=q\eta\sin\phi ,
\label{polar}
\end{eqnarray}
with
\begin{eqnarray}
p=\sqrt{d/d_{2}},\ \ q=\sqrt{d/d_{3}}, \ \
\  d=\frac{1}{2}(d_{2}+d_{3})\, .
\label{pqd}
\end{eqnarray}
An analytical solution for the spectrum of $H_{\mbox{\scriptsize qo}}$ can be
found for a specific set of parameters, when assuming coherent QO
oscillations (the so-called coherent QO mode, CQOM) with a frequency
$\omega=\sqrt{C_2/B_2}=\sqrt{C_3/B_3}\equiv \sqrt{C/B}$.  Then, the
two-dimensional potential in Hamiltonian (\ref{HqoK}) obtains a shape with an
ellipsoidal equipotential bottom in the space of the collective deformation
variables $\tilde{\beta_2}$ and $\tilde{\beta_3}$  \cite{b2b3mod}. This
allows a separation of the ellipsoidal variables and reduction of the problem
to the one-dimensional Schr\"odinger equation for an analytically solvable
potential of Davidson type in the radial variable $\eta$. The motion with
respect to this potential corresponds to a ``soft'' QO vibration mode without
fixed minima in $\tilde{\beta_2}$ and $\tilde{\beta_3}$. These should not be
confused with the fixed $\beta_2$- and $\beta_3$- deformations in the s.p.
Woods-Saxon potential of the DSM. The CQOM approach has been successfully
applied to QO spectra of even-even and odd mass nuclei
\cite{b2b3mod,b2b3odd,MDSSL12,MDDSLS13}.

The quadrupole and octupole semiaxes $\tilde{\beta}_{2}^{\mbox{\scriptsize
sa}}$ and $\tilde{\beta}_{3}^{\mbox{\scriptsize sa}}$ of the ellipsoidal CQOM
potential bottom are defined  for even-even nuclei  as \cite{MDSSL12}
\begin{eqnarray}
\tilde{\beta}_{\lambda}^{\mbox{\scriptsize sa}}(I)
=[2X(I)/d_{\lambda}C_{\lambda}]^{1/4},
\qquad \lambda=2,3,
\label{semiax}
\end{eqnarray}
with the centrifugal factor $X(I)=[d_0+I(I+1)]/2$. For an odd-$A$ nucleus the
expression for the semixes takes the form
\begin{eqnarray}
\tilde{\beta}_{\lambda}^{\mbox{\scriptsize sa}}(I^{\pi},K_{b})
=[2\widetilde{X}(I^{\pi},K_{b})/d_{\lambda}C_{\lambda}]^{1/4},
\qquad \lambda=2,3,
\label{semiaxtilde}
\end{eqnarray}
with $\widetilde{X}(I^{\pi},K_{b})$ determined in Eq.~(\ref{Xmix}). Comparing
the expressions of $X(I)$ and $\widetilde{X}(I^{\pi},K_{b})$ it becomes clear
that for  the odd-nucleus the semiaxes $\tilde{\beta}_{\lambda
=2,3}^{\mbox{\scriptsize sa}}(I^{\pi},K_{b})$ in Eq.~(\ref{semiaxtilde})
differ from the semiaxes $\tilde{\beta}_{\lambda}^{\mbox{\scriptsize sa}}(I)$
(\ref{semiax}) of the original even-even core CQOM potential because here the
$\widetilde{X}(I^{\pi},K_{b})$ factor includes the additional term
$(-K_{b}^2)$ as well as the Coriolis mixing and decoupling contributions from
the single nucleon. The CQOM potential semiaxes obey the relation
\cite{MDSSL12}
\begin{eqnarray}
\frac{\tilde{\beta}_{3}^{\mbox{\scriptsize sa}}}
{\tilde{\beta}_{2}^{\mbox{\scriptsize sa}}}
=\frac{1}{\sqrt{2p^{2}-1}},
\label{semrat}
\end{eqnarray}
with $p$ defined in Eq.~(\ref{pqd}) determining the relative contribution of
the quadrupole and octupole collective modes in the coherent QO motion. We
note that the value $p=1$ corresponds to equal values of both semiaxes, i.e.,
to a circle form of the CQOM potential bottom. In terms of the coherence
assumption concept this means that both the quadrupole $\tilde{\beta_2}$ and
octupole $\tilde{\beta_3}$ deformation modes enter the collective CQOM motion
with the same weight. This case will be discussed later in the paper and is
exemplified in Figs.~\ref{wfdensities} and \ref{wfdenscont15div2}.

Despite the missing single ($\tilde{\beta_{2}},\tilde{\beta}_{3}$)- minimum
in the CQOM potential, the collective QO states of the system are still
characterized by the so-called \emph{dynamical} deformations determined by
the density maxima of the QO vibration wave function. Explicitly, the CQOM QO
vibration wave function is given by \cite{b2b3mod,MDSSL12}
\begin{eqnarray}
\Phi^{\pi_{\mbox{\scriptsize qo}}}_{nkI} (\eta,\phi)=
\psi_{nk}^{I}(\eta )\varphi^{\pi_{\mbox{\scriptsize qo}}}_{k}(\phi),
\label{wvib}
\end{eqnarray}
where the radial part
\begin{equation}
\psi^I_{nk}(\eta )=\sqrt
{\frac {2c\Gamma(n+1)}{\Gamma(n+2s+1)}}
e^{-c\eta^2/2}(c\eta^{2})^sL^{2s}_n(c\eta^2)\
\label{psieta1}
\end{equation}
involves generalized Laguerre polynomials in the variable $\eta$, with
$s=(1/2)\sqrt{k^2+b\widetilde{X}(I,K)}$ and $c=\sqrt{BC}/\hbar$, the latter
having the meaning of a reduced QO oscillator frequency, and $\Gamma (z)$
denotes the Gamma function. The angular part in the variable $\phi$ appears
with a positive or negative parity $\pi_{\mbox{\scriptsize qo}}$ of the
collective QO mode as follows
\begin{eqnarray}
\varphi_{k}^{+}(\phi)&=& \sqrt{2/\pi}\cos (k\phi ) \ , \qquad k=1, 3, 5,
...\ ;\label{parplus} \\
\varphi_{k}^{-}(\phi)&=& \sqrt{2/\pi}\sin (k\phi ) \ , \qquad k=2, 4, 6,
...\ . \label{parminus}
\end{eqnarray}

The maxima of the density $|\Phi^{\pi_{\mbox{\scriptsize
qo}}}_{nkI}|^{2}\rightarrow |\Phi^{\pi_{\mbox{\scriptsize
qo}}}_{nkI}(\tilde{\beta_2},\tilde{\beta_3})|^{2}$ calculated in the
$(\tilde{\beta_2},\tilde{\beta_3})$ space pin down the dynamical deformation
values \cite{MDSSL12}. Strictly speaking, the dynamical deformation is
defined by the expectation value of the square of the corresponding multipole
(deformation) operator in the CQOM state, but considering the density maximum
is enough to locate its position in the $(\tilde{\beta_2},\tilde{\beta_3})$
space. The positions of these maxima  are situated outside of the potential
bottom ellipse and move  further out with increasing angular momentum. They
essentially characterize the collective dynamical behaviour of the nucleus in
the presence of a coherent mode. This will be illustrated in
Sec.~\ref{cqomfits} for the present model application in $^{229}$Th. It will
be seen that the CQOM dynamical deformations appearing in the overall
collective spectrum of the nucleus are reasonably correlated with the
intrinsic Woods-Saxon DSM QO deformations.

We should stress here, however, that the dynamical QO deformations in CQOM do
not need to ultimately coincide or even to be close to the fixed Woods-Saxon
deformations $\beta_2$ and $\beta_3$ of the DSM. Imposing artificially such a
constraint  would deprive the overall algorithm of the capability to
incorporate the individual (separate) dynamic properties of the collective
and s.p. degrees of freedom (carried by the available data) and, therefore,
of the possibility to plausibly reproduce the interaction between them.  The
present model formalism does not put a constrain on both potentials but
rather leaves them to independently feel, as much as possible, the
corresponding physical conditions which govern the nuclear collective and
intrinsic motions and their very fine interplay. As it will be seen in the
following Sec.~\ref{modappl}, in the case of $^{229m}$Th, the DSM
deformations $\beta_2$ and $\beta_3$ determine the hyperfine (from the
nuclear point of view) conditions for the appearance of the $K^{\pi}=3/2^{+}$
isomer while the dynamical CQOM deformations in the
$(\tilde{\beta_2},\tilde{\beta_3})$-space reflect the conditions imposed by
the overall collective spectrum which complement the microscopic
isomer-formation mechanism.

By taking the analytical CQOM solution together with the result of the DSM
plus BCS calculation the QO core plus particle spectrum built on the given
q.p. bandhead state is obtained in the form \cite{Minkov_Palffy_PRL_2017}
\begin{equation}
E_{nk}^{\mbox{\scriptsize tot}}(I^{\pi} ,K_{b})
=\epsilon^{K_{b}}_{\mbox{\scriptsize qp}}
+ \hbar\omega \left[ 2n+1+\sqrt{k^2+b\widetilde{X}(I^{\pi},K_{b})}\right].
\label{enspect1}
\end{equation}
Here $b=2B/(\hbar^2 d)$ has the meaning of a reduced inertia parameter, while
$n=0,1,2,...$ and $k=1,2,3,...$ stand for the radial and angular QO
oscillation quantum numbers, respectively, with $k$ odd (even) for the
even-parity (odd-parity) states of the core \cite{b2b3mod,b2b3odd}. The
levels of the total QO core plus particle system, determined by the given $n$
and pair of $k^{(+)}$ and $k^{(-)}$ values for the states with $I^{\pi=+}$
and $I^{\pi=-}$, respectively, form a split doublet with respect to the
parity, called a quasi-parity-doublet \cite{b2b3odd,MDDSLS13}.

The corresponding wave functions can be constructed in three steps. First,
the quadrupole-octupole vibration wave function of the CQOM is calculated
according to Eq.~(\ref{wvib}). Second, we can construct the unperturbed QO
core plus particle wave function \cite{NM13,MDDSLS13}:
\begin{equation}
\begin{aligned}
&\Psi^{\pi ,\pi^{b}}_{nkIMK}(\eta ,\phi ,\theta)=
\frac{1}{N_{K}^{(\pi^{b})}}\sqrt{\frac{2I+1}{16\pi^2}}
\Phi^{\pi\pi^{b}}_{n k I} (\eta,\phi)\\
\times&\left[ D^{I}_{M\, K}(\theta )\mathcal{F}^{(\pi^{b})}_K+
\pi\pi^{b}(-1)^{I+K}D^{I}_{M\,-K}(\theta)\mathcal{F}^{(\pi^{b})}_{-K}\right]\, ,
\end{aligned}
\label{wfpcore}
\end{equation}
where $D^{I}_{M\, K}(\theta )$ are the rotation (Wigner) functions and
$\Phi^{\pi\pi^{b}}_{n k I} (\eta,\phi)$ are the QO vibration functions
(\ref{wvib}) with $\pi_{\mbox{\scriptsize qo}}=\pi\pi^{b}$. In
Eq.~(\ref{wfpcore}) the relevant part
$\mathcal{F}^{(\pi^{b})}_K=\mathcal{F}^{(+)}_K$ or $\mathcal{F}^{(-)}_K$ of
the s.p. wave function $\mathcal{F}_K$ given in Eq.~(\ref{ahodecomp}) is
taken by projecting the latter with respect to the experimentally assigned
bandhead parity $\pi^{b}=+$ or $-$, thus providing a good parity of the total
core-plus-particle wave function.

Finally, the Coriolis perturbed wave function
$\widetilde{\Psi}\equiv\widetilde{\Psi}^{\pi ,\pi^{b}}_{nkIMK_{b}}$
corresponding to Hamiltonian (\ref{Htotal}) with the spectrum
(\ref{enspect1}) is obtained in the first order of perturbation theory and
has the form
\begin{equation}
\widetilde{\Psi}=\frac{1}{\widetilde{N}_{I\pi K_{b}}}
\left[\Psi^{\pi ,\pi^{b}}_{nkIMK_{b}} + A\sum_{\nu \neq b}
C^{I\pi}_{K_{\nu} K_{b}}\Psi^{\pi ,\pi^{b}}_{nkIMK_{\nu}}\right],
\label{wtcoriol}
\end{equation}
where $K_{\nu}= K_{b}\pm 1,\frac{1}{2}$, the expansion coefficients read
\begin{equation}
C^{I\pi}_{K_{\nu} K_{b}}=\frac{\widetilde{a}_{K_{\nu}K_{b}}^{(\pi ,\pi^{b})}(I)}
{\epsilon^{K_{\nu}}_{\mbox{\scriptsize qp}}-
\epsilon^{K_{b}}_{\mbox{\scriptsize qp}}},
\end{equation}
while the normalization factor is given by
\begin{equation}
\begin{aligned}
&\widetilde{N}_{I\pi K_{b}}^{2}=\left\langle\widetilde{\Psi}^{\pi
,\pi^{b}}_{nkIMK_{b}}\big| \widetilde{\Psi}^{\pi
,\pi^{b}}_{nkIMK_{b}}\right\rangle   \\
&=1+2A\mkern-18mu\!\!\sum_{\substack {\nu \neq b\\K_{\nu}= K_{b}=\frac{1}{2}}}
\mkern-18mu\!\!C^{I\pi}_{K_{\nu} K_{b}}\delta_{K_{\nu}K_{b}}
\frac{P^{b}_{K_{\nu}K_{b}}}{N^{(\pi^{b})}_{K_{\nu}}N^{(\pi^{b})}_{K_{b}}}
\left\langle{\mathcal{F}_{K_{\nu}}}^{(\pi^{b})}\big|
{\mathcal{F}_{K_{b}}}^{(\pi^{b})}\right\rangle   \\
&+ A^{2}\mkern-18mu\!\!\sum_{\substack {\nu_{1,2} \neq b\\
 K_{\nu_{1},\nu_{2}}=K_{b}\pm 1,\frac{1}{2} }}
\!\!\mkern-18mu\
C^{I\pi}_{K_{\nu_{1}} K_{b}}C^{I\pi}_{K_{\nu_{2}} K_{b}}\delta_{K_{\nu_{1}}K_{\nu_{2}}}
\\
&\times \frac{P^{b}_{K_{\nu_{1}}K_{\nu_{2}}}}
{N^{(\pi^{b})}_{K_{\nu_{1}}}N^{(\pi^{b})}_{K_{\nu_{2}}}}
\left\langle{\mathcal{F}_{K_{\nu_{1}}}}^{(\pi^{b})}\big|
{\mathcal{F}_{K_{\nu_{2}}}}^{(\pi^{b})}\right\rangle .
\end{aligned}
\label{psicornorm}
\end{equation}

\subsection{Electric and magnetic transition rates}
\label{transit}
Expressions for the reduced $B(E1)$-, $B(E2)$- and $B(E3)$- probabilities for
transitions between states with energies given by Eq.~(\ref{enspect1}) and
Coriolis perturbed wave function given by Eq.~(\ref{wtcoriol}) are derived by
using the electric transition operators in the general form
\begin{eqnarray}
 Q_{\mu}(E\lambda)&=&\sqrt\frac{2\lambda +1}{4\pi (4-3\delta
_{\lambda,1})}\hat{Q}_{\lambda 0}\sum_{\nu}D^{\lambda}_{\mu\nu}, \nonumber\\
\lambda &=&1,2,3,\ \ \mu=0,\pm 1, ..., \pm\lambda ,
\label{melambda}
\end{eqnarray}
with the explicit form of the operators $\hat{Q}_{\lambda 0}$ given by Eqs.
(31)--(33) in \cite{MDSSL12}.

The expression for the $B(M1)$ reduced transition probability was obtained by
using the standard core plus particle magnetic dipole ($M1$) operator (e.g.
see Eq.~(3.61) in Ref.~\cite{Ring1980}) written as
\begin{equation}
\hat{M}1=\sqrt{\frac{3}{4\pi}}\mu_{N}\left [ g_{R}(\hat{I}-\hat{j})
+g_s\, \hat{s}+g_l\, \hat{l} \right ],
\label{m1operator}
\end{equation}
after taking it in the intrinsic frame. The operators $\hat{s}$ and $\hat{l}$
in Eq.~(\ref{m1operator}) correspond to the s.p. spin and orbital momenta and
$\hat{j}=\hat{l}+\hat{s}$. The quantities $g_s$ and $g_l$ are the spin and
orbital gyromagnetic factors, respectively, and $g_R$ is the collective
gyromagnetic factor. The orbital factor is $g_l=0\, (1)$ for neutrons
(protons), while the spin factor is taken as $g_s=q_s\ g_s^{\mbox{\scriptsize
free}}$, with $g_s^{\mbox{\scriptsize free}}=-3.826\, (5.586)$ for neutrons
(protons) \cite{Ring1980}. The quantity $q_s$ is an attenuation factor
usually supposed to be $q_s=0.6-0.7$, taking into account spin-polarization
effects \cite{Mottelson60}. The collective gyromagnetic factor $g_R$ is often
associated with the ratio $g_R=Z/(Z+N)$, with $Z$ and $N$ being the proton
and neutron numbers, respectively, adopted  on the basis of the
liquid-drop-model \cite{Way39}. However, it is known that in most deformed
nuclei  $g_R$ is lowered with respect to this ratio by 20{\%}-30{\%} or more
\cite{Boden62,EG70}, with the attenuation being explained by the influence of
the pairing interaction on the collective moment of inertia
\cite{NP61,PBN68,Greiner65}. Therefore, in Ref.~\cite{Minkov_Palffy_PRL_2019}
we have introduced the relevant quenching factor $q_R$ such that $g_R=q_R
Z/(Z+N)$, showing that on the basis of several earlier theoretical and
experimental analyses, it can be taken for $^{229}$Th as low as $q_R\sim
0.6$. Below it will be seen that both attenuation factors $q_s$ and $q_R$
play an important role in the model prediction of the $B(M1)$ transition
rates and magnetic moments and their consideration with further slightly
lower values may shed more light on the $^{229}$Th formation mechanism.

The following common form of the expressions for both types ($T$) of the
electric ($T=E$) and magnetic ($T=M$) transition with multipolarity $\lambda$
between initial (i) and final (f) states was derived
\cite{Minkov_Palffy_PRL_2017}
\begin{equation}
\label{emcormix}
\begin{aligned}
&B(T\lambda; \pi^{b_i} I_{i}\pi_i K_i\rightarrow \pi^{b_f} I_{f}\pi_f K_f) \\
&= R^{T\lambda}
\delta_{\pi^{b_f} \pi^{b_i}}\left[(1+\pi_f\pi_i(-1)^{\lambda \delta_{T,E}})/2 \right] \\
&\times \frac{1}{\widetilde{N}_{I_{f}\pi_{f}K_f}^{2}\widetilde{N}_{I_{i}\pi_{i}K_i}^{2}}
\Bigg[\delta_{K_fK_i}C^{I_fK_f}_{I_iK_i\lambda0}
\frac{P^{b_fb_i}_{K_{f}K_{i}} M^{\pi^{b_f}\pi^{b_i}}_{K_{f}K_{i}}}
{N_{K_f}^{(\pi^{b_f})}N_{K_i}^{(\pi^{b_i})}}\Bigg. \\
&+ A\, \Bigg. C^{I_fK_f}_{I_iK_f\lambda0}\mkern-18mu\!\!\sum_{\substack{\nu\neq i
\\ K_{\nu}=K_i\pm 1,\frac{1}{2}}}\mkern-18mu\!\!\delta_{K_fK_{\nu}}C^{I_{i}\pi_{i}}_{K_{\nu}K_i}
\frac{P^{b_f}_{K_{f}K_{\nu}} M^{\pi^{b_f}\pi^{b_i}}_{K_{f}K_{\nu}}}
{N_{K_f}^{(\pi^{b_f})}N_{K_{\nu}}^{(\pi^{b_i})}}\Bigg. \\
&+ \Bigg. A\, C^{I_fK_i}_{I_iK_i\lambda0}\mkern-2mu\!\!\sum_{\substack{\nu\neq f
\\K_{\nu}=K_f\pm 1,\frac{1}{2} }}\mkern-18mu\!\!\delta_{K_{\nu}K_i}C^{I_{f}\pi_{f}}_{K_{\nu}K_f}
\frac{P^{b_i}_{K_{\nu}K_{i}} M^{\pi^{b_f}\pi^{b_i}}_{K_{\nu}K_{i}}}
{N_{K_{\nu}}^{(\pi^{b_f})}N_{K_{i}}^{(\pi^{b_i})}}\Bigg.  \\
&+ A^{2}\Bigg. \mkern-20mu\!\!\sum_{\substack{\nu''\neq  f
\\ K_{\nu''}=K_f\pm 1,\frac{1}{2}}} \sum_{\substack{\nu'\neq i
\\ K_{\nu'}=K_i\pm 1,\frac{1}{2}}}\mkern-20mu\!\! \delta_{K_{\nu''}K_{\nu'}}
C^{I_fK_{\nu''}}_{I_iK_{\nu'}\lambda K_{\nu''}-K_{\nu'}} \\
&\times
C^{I_{f}\pi_{f}}_{K_{\nu''}K_f} C^{I_{i}\pi_{i}}_{K_{\nu'}K_i}
\frac{P^{b_fb_i}_{K_{\nu''}K_{\nu'}}M^{\pi^{b_f}\pi^{b_i}}_{K_{\nu''}K_{\nu'}}}
{N_{K_{\nu''}}^{(\pi^{b_f})}N_{K_{\nu'}}^{(\pi^{b_i})}} \Bigg]^{2},
\end{aligned}
\end{equation}
where the factor
\begin{equation}
R^{T\lambda=E\lambda}=\frac{2\lambda +1}{4\pi (4-3\delta_{\lambda,1})}
R_{\lambda}^{2}( \pi^{b_i} n_i k_i I_{i}\rightarrow \pi^{b_f} n_f
k_f I_{f})
\label{transinteg}
\end{equation}
involves integrals on the radial and angular variables in CQOM (see
Eqs.~(35)--(41) and Appendixes B and C in Ref.~\cite{MDSSL12}) and
\begin{equation}
R^{T1=M1}=\frac{3}{4\pi}\mu_{N}^{2}
\end{equation}
involves the nuclear magneton $\mu_N$. Also here
\begin{equation}
M^{\pi^{b_f}\pi^{b_i}}_{K_{f}K_{i}}
=\left\{\mkern-2mu\!\!\begin{array}{ll} \langle\mathcal{F}_{K_{f}}^{(\pi^{b_f})}|
\mathcal{F}_{K_{i}}^{(\pi^{b_i})}\rangle ,\hspace{2.4cm} \mbox{for}\ T=E & \ \ \  \\
\left[(g_l-g_R)K_{i}\delta_{K_fK_i}\langle\mathcal{F}_{K_{f}}^{(\pi^{b_f})}|
\mathcal{F}_{K_{i}}^{(\pi^{b_i})}\rangle \right. &  \\
\left.  +(g_s-g_l)\langle\mathcal{F}_{K_{f}}^{(\pi^{b_f})}|\hat{s}_{0}|
\mathcal{F}_{K_{i}}^{(\pi^{b_i})}\rangle\right], \ \mbox{for}\ T=M,& \ \ \
\end{array} \right.
\label{Mfactors}
\end{equation}
where $\hat{s}_{0}$ is the $z$ component of the spin operator in spherical
representation. The factors $C^{I_2K_2}_{I_1K_1\lambda \mu}$ in
Eq.~(\ref{emcormix}) are Clebsch-Gordan coefficients. The integrals in
Eq.~(\ref{transinteg}) depend on the model parameters $c$, defined below
Eq.~(\ref{psieta1}), and $p$, Eq.~(\ref{pqd}), both determining the electric
transition probabilities \cite{MDSSL12}.

The reduced transition probability expression (\ref{emcormix}) contains
first-order and second-order $K$-mixing effects. First-order mixing terms
practically contribute with nonzero values only in the cases
$K_{i/f}=K_{\nu}=1/2$, i.e., when a $K_{i/f}=1/2$ bandhead state is mixed
with another $K_{\nu}=1/2$ state present in the considered range of admixing
orbitals. A second-order mixing effect connects states with $\Delta K=1,2$
and allows different combinations of $|K_i-K_f|\leq 2$ which provide
respective nonzero contribution of the Coriolis mixing to the transition
probability. In this way the present formalism provides nonzero transition
probabilities between states with different $K$-values despite the axial
symmetry assumed in both CQOM and DSM parts of Hamiltonian (\ref{Htotal}). We
stress that although often disregarded in the literature, it is only through
the Coriolis mixing that the $M1$ and $E2$ isomer decay channels for
$^{229m}$Th are rendered possible within the model discussed here.

\subsection{Magnetic moment}
\label{magmom}
The  described model formalism allows us to obtain the magnetic-dipole moment
in any state of the quasiparity doublet spectrum characterized by the
Coriolis perturbed wave function $\widetilde{\Psi}_{IMK_{b}}$
(\ref{wtcoriol}).  The magnetic moment is determined by the matrix element
$\mu = \sqrt{\frac{4\pi}{3}}\langle \widetilde{\Psi}_{IIK_{b}}|\hat{M}1_{0}
|\widetilde{\Psi}_{IIK_{b}}\rangle$, where $\hat{M}1_{0}$ is the zeroth
spherical tensor component of the operator $\hat{M}1$,
Eq.~(\ref{m1operator}), taken after transformation into the intrinsic frame
(see Chapter 9 of Ref.~\cite{EG70}). Thus we obtain the following expression
for the magnetic moment in a state with collective angular momentum $I$ and
parity $\pi$ built on a q.p. bandhead state with $K=K_{b}$ and $\pi
=\pi^{b}$:
\begin{equation}
\begin{aligned}
\hspace{-0.45cm}
&\mu=\mu_{N}g_{R}I+\frac{1}{I+1}\frac{1}{\widetilde{N}_{I\pi K_{b}}^{2}}
\Bigg[K_{b}\frac{M^{\pi^{b}}_{K_{b}K_{b}}}{N^{(\pi^{b})}_{K_{b}}} \\
&+ 2A\, K_{b}\sum_{\substack {\nu \neq b\\   K_{\nu}= K_{b}=\frac{1}{2}}}
\delta_{K_{\nu}K_{b}}C^{I\pi}_{K_{\nu} K_{b}}
\frac{P^{b}_{K_{\nu}K_{b}} M^{\pi^{b}}_{K_{\nu}K_{b}}}
{N^{(\pi^{b})}_{K_{\nu}}N^{(\pi^{b})}_{K_{b}}} \\
& + A^{2}\mkern-18mu\!\!\sum_{\substack {\nu_{1,2} \neq b\\
 K_{\nu_{1},\nu_{2}}=K_{\nu} \\ =K_{b}\pm 1,\frac{1}{2}}}
\!\!\mkern-18mu\delta_{K_{\nu_{1}}\! K_{\nu_{2}}}\!K_{\nu}
C^{I\pi}_{K_{\nu_{1}}\! K_{b}}C^{I\pi}_{K_{\nu_{2}}\!K_{b}}
\frac{P^{b}_{K_{\nu_{1}}\!K_{\nu_{2}}}M^{\pi^{b}}_{K_{\nu_{2}}\!K_{\nu_{1}}}}
{N^{(\pi^{b})}_{K_{\nu_{1}}}N^{(\pi^{b})}_{K_{\nu_{2}}}}
\Bigg],
\end{aligned}
\label{mumod}
\end{equation}
with $M^{\pi^{b}}_{K_{\mu}K_{\nu}}\equiv
M^{\pi^{b},\pi^{b}}_{K_{\mu}K_{\nu}}$ being defined in Eq.~(\ref{Mfactors})
($T=M$) and all other quantities being already defined above. We note that
the complete expression would involve an additional decoupling term applying
for the case of $K_{b}=1/2$ appearing after transforming the $\hat{M}1$
operator (\ref{m1operator}) into the intrinsic frame \cite{EG70}. Here we do
not take it into account in Eq.~(\ref{mumod}), since in the present
application of the model to $^{229}$Th no $K=1/2$ bandheads appear. The
second and the third term in the brackets of Eq.~(\ref{mumod}) take into
account the influence of the Coriolis mixing on the magnetic moment. In fact,
also the second term only applies for $K_{b}=1/2$, but we keep it for
consistency with the $B(E\lambda)$ and $B(M1)$ expressions (\ref{emcormix}).
Thus, in the present application of the model only the third term is
important for the Coriolis mixing in the magnetic moment.

One can easily check that in the case of missing Coriolis mixing
Eq.~(\ref{mumod}) appears in the usual form of the particle-rotor expression,
e.g. Eq.~(3.62) in Ref.~\cite{Ring1980}, in which the intrinsic gyromagnetic
ratio $g_{K}$ is
\begin{eqnarray}
g_{K_{b}}=\frac{1}{K_{b}} \frac{1}{[N_{K_{b}}^{(\pi^{b})}]^2}\langle
\mathcal{F}^{(\pi^{b})}_{K_{b}}|g_s\cdot\Sigma +g_l\cdot\Lambda |
\mathcal{F}^{(\pi^{b})}_{K_{b}} \rangle .
\label{gKb}
\end{eqnarray}
Equation~(\ref{gKb}) still takes into account the circumstance that in the
case of nonzero octupole deformation we have to apply the projected and
renormalized s.p. wave function as explained below Eq.~(\ref{wfpcore}). In
the case of missing octupole deformation (nonmixed s.p. wave function),
Eq.~(\ref{gKb}) reduces to the standard ``reflection-symmetric'' expression
(3.63) in Ref.~\cite{Ring1980}. In this way the present model expression for
the magnetic-dipole moment in Eq.~(\ref{mumod}) is consistent with the
relevant limiting cases.

\subsection{Model application in $^{229}$Th}
\label{modappl}

The CQOM plus DSM-BCS model framework described above contains a number of
parameters that are determined according to the physical conditions which
govern the structure and dynamics of the nucleus $^{229}$Th and to the
available experimental data. These parameters are the two already discussed
Woods-Saxon DSM QO deformations $\beta_2$ and $\beta_3$, the five CQOM
parameters, namely the QO oscillator frequency $\omega$, the reduced inertia
factor $b$ in Eq.~(\ref{enspect1}), the parameter $d_0$ in Eq.~(\ref{Xmix})
and the parameters $c$ and $p$ from Eqs.~(\ref{psieta1}) and (\ref{pqd}),
respectively, entering Eq.~(\ref{transinteg}), the Coriolis mixing constant
$A$, and the two pairing parameters $g_0$ and $g_1$ entering Eq.~(\ref{Gnp}).
As it will be detailed below, the first two (Woods-Saxon DSM QO deformation)
parameters are determined in a region of the deformation space providing
ultimate DSM conditions for the formation of the $^{229m}$Th isomer.  The
pairing constants are fixed for the overall study to values in a range
typical for the adjacent regions of nuclei. Finally, the five CQOM parameters
and the Coriolis mixing constant are adjusted in a fitting procedure to
quantitatively reproduce the positive- and negative-parity levels of
$^{229}$Th with energy below 400 keV as well as the available experimental
data on transition rates and magnetic moments at each particular Woods-Saxon
DSM QO deformation. As explained in Sec.~\ref{wf}, the five CQOM parameters
also determine the shape of the collective potential and the corresponding
dynamical deformations in the ($\tilde{\beta}_{2},\tilde{\beta}_{3}$)-space.
The rather fine parameter determination procedure described here is based on
the following physical assumptions:

\begin{enumerate}
\item The considered part of the spectrum consists of two
    quasi-parity-doublets: an yrast one, based on the $K_{b}=5/2^{+}$ GS
    corresponding to the 5/2[633] s.p. orbital and a nonyrast
    quasi-parity-doublet, built on the isomeric $K_{b}=3/2^{+}$ state
    corresponding to the 3/2[631] orbital. Both orbitals are very close to
    each other providing a quasidegeneracy of the GS and IS. This condition
    primary depends on the choice of the quadrupole ($\beta_2$) and
    octupole ($\beta_3$) deformation parameters in DSM and on the BCS
    pairing contribution in the q.p. energy of both states.

\item Both quasi-parity-doublets correspond to coherent QO vibrations and
    rotations with the same radial-oscillation quantum number $n=0$, the
    lowest possible angular-oscillation number $k^{(+)}=1$ for the
    positive-parity sequences and one of the few lowest possible
    $k^{(-)}=2,4,6$ values for the negative-parity states [see
    Eq.~(\ref{enspect1}) and the text below it]. Hereinafter we consider
    only the lowest $k^{(-)}=2$ value in the both quasi-parity-doublets.
    This suggests completely identical QO vibration modes superposed on
    both GS and IS. The vibration modes alone obviously do not cause any
    mutual displacement of the two quasi-parity-doublets, but the term
    $K^{2}_{b}$ in the centrifugal expression
    $\widetilde{X}(I^{\pi},K_{b})$ in  Eq.~(\ref{Xmix}) does. It directly
    mixes the collective energy with the bandhead and down-shifts the
    $K_{b}=5/2^{+}$ level sequences with respect to the $K_{b}=3/2^{+}$
    ones. This term affects the mutual displacement of IS and GS and,
    therefore, plays a role in the finally observed quasidegeneracy effect.

\item The Coriolis mixing affects the total spectrum and the IS-GS
    displacement as well through the corresponding perturbation sum in
    Eq.~(\ref{Xmix}). As realized in Ref.~\cite{Minkov_Palffy_PRL_2019},
    the mixing directly affects the $I_{b},K_{b}=5/2^{+}$ GS which gets an
    admixture from the $I=5/2^{+}$ state of the IS-based band, whereas the
    $I_{b},K_{b}=3/2^{+}$ IS remains unmixed due to the missing $I=3/2^{+}$
    counterpart in the yrast (GS) band. The corresponding effect of the
    Coriolis mixing in the GS is that it lowers the value of the GS
    magnetic moment. On the other hand it raises the $B(M1)$ and $B(E2)$
    transition probabilities.

\end{enumerate}

\begin{figure*}
\centering
\includegraphics[width=8cm]{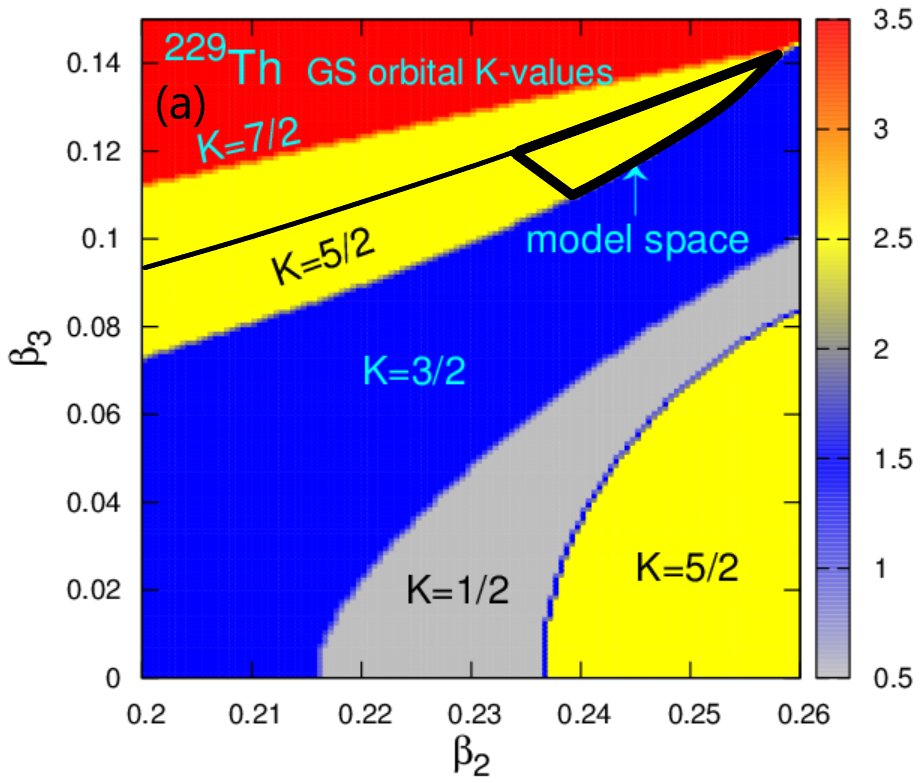}
\includegraphics[width=8cm]{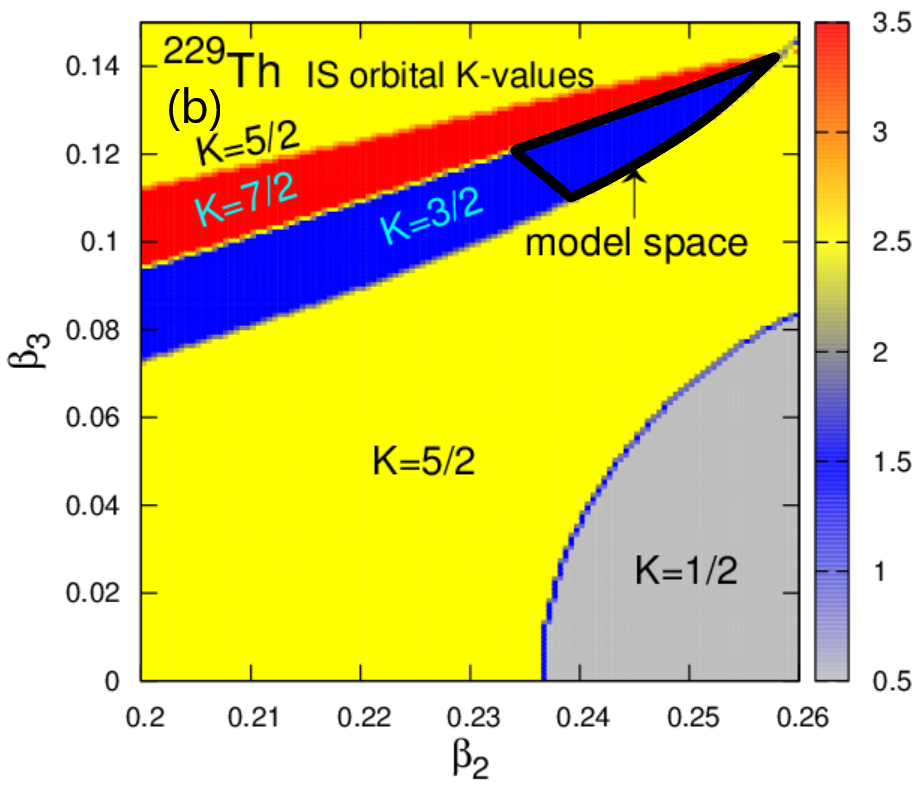}
\includegraphics[width=8cm]{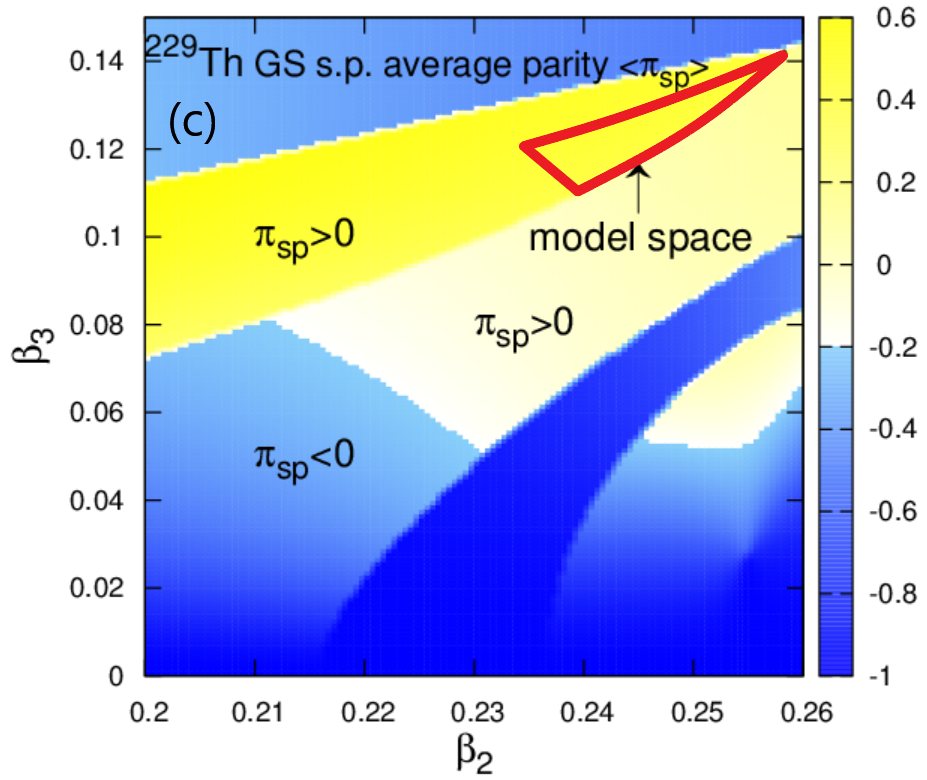}
\includegraphics[width=8cm]{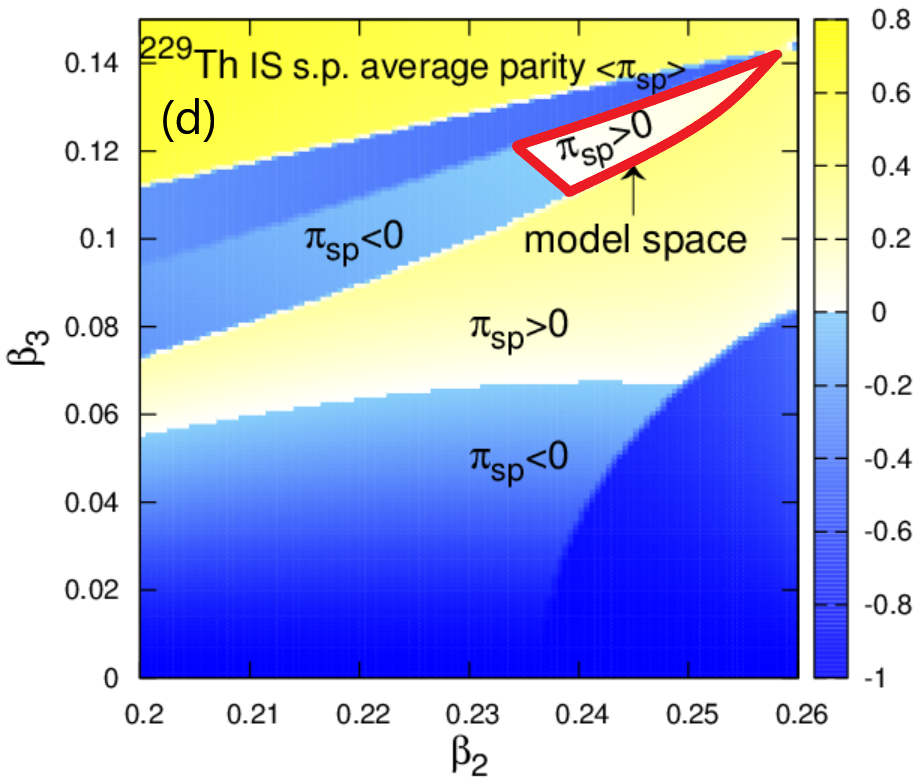}
\caption{$K$-values for the (a) GS and (b) IS s.p. orbitals and the respective
average parities (c), (d) $\langle\pi_{\mbox{\scriptsize sp}}\rangle$ appearing
in the DSM within the space of quadrupole and octupole deformations. The regions
of relevant deformations providing the correct $K_{GS}=5/2$ and $K_{IS}=3/2$
values with $\langle\pi_{\mbox{\scriptsize sp}}\rangle >0$ are delimitated by
thick contour lines. The thinner black line in plot (a) maps the relevant border
of the $K_{IS}=3/2$ region from plot (b).}
\label{th229_kpi}
\end{figure*}


The assumptions above sketch the mechanism which may lead to the formation of
a quasidegenerate pair of $5/2^{+}$ GS and $3/2^{+}$ IS in $^{229}$Th. We see
that the very fine interplay between the involved collective and s.p. degrees
of freedom is directly governed by the Woods-Saxon DSM QO deformations
$\beta_2$ and $\beta_3$, the pairing strength determined by the parameters
$g_0$ and $g_1$ in Eq.~(\ref{Gnp}) and the Coriolis mixing strength
determined by the parameter $A$ in (\ref{Xmix}). The remaining CQOM
parameters $\omega$, $b$, $d_0$, $c$ and $p$ influence the isomer energy
through the overall fit of the energy spectrum, transition rates and magnetic
moments. Within the above physical mechanism the $3/2^{+}$ IS of $^{229}$Th
appears as an essentially s.p., i.e., microscopic, effect, the energy and
electromagnetic properties of which, however, are formed under the influence
of the collective dynamics of the nucleus.

In Ref.~\cite{Minkov_Palffy_PRL_2017} the above algorithm was applied through
several steps, including the choice of $\beta_2$ and $\beta_3$ in DSM based
on information available for neighbouring even-even nuclei (see the beginning
of next section), tuning of the pairing constants in BCS to reach a rough
proximity of GS and IS and subsequent fine adjustment of the collective CQOM
parameters together with the $K$-mixing constant $A$ to obtain overall model
description and predictions. It was demonstrated that at the expense of a
minor deterioration of the agreement between the overall theoretical and
experimental spectrum, one can exactly reproduce the IS energy of about 8 eV.
Of course, such a refinement is of  little practical significance since it is
beyond the genuine accuracy provided by any nuclear structure model.

Few comments regarding the results in the next section should be given here
in advance. We remark that some model parameters are not completely
independent regarding particular physical observables. Thus, the change in
the IS-GS displacement due to variation in the DSM QO deformations could be
compensated by variations in the pairing constants or the $K$-mixing constant
$A$. Therefore, one of the important issues to be clarified is the extent to
which the different model parameters are correlated in the problem and how we
can constrain them to reach most unambiguously the correct solution. Our
numerical study showed that if we fix the pairing parameters in
Eq.~(\ref{Gnp}) to the values of $g_0=18.805$ MeV and $g_1=7.389$ MeV, which
were  tuned in the model description in Ref.~\cite{Minkov_Palffy_PRL_2017},
the further analysis and drawn conclusions also apply for the pairing
strengths adopted in Refs. \cite{NR1995} and \cite{WM10}. Therefore,
hereinafter we use the above fixed $g_0$ and $g_1$ parameter values while
directing our study to the examination of the QO deformation space of the
DSM. Another point is that in Ref.~\cite{Minkov_Palffy_PRL_2019} the IS and
GS magnetic moments were predicted without taking their experimental values
into the model adjustment procedure.  In the present work  we include the
magnetic moments into the fitting procedure by considering all observables in
the fit analysis (energies, transition rates and magnetic moments) on the
same footing. We will also investigate to what extent the gyromagnetic
quenching factors $q_s$ and $q_R$ can be reasonably varied for the
reproduction of the GS and IS magnetic moments. This analysis aims to reduce
the arbitrariness in the model predictions for the $^{229}$Th IS properties.

\begin{figure*}
\centering
\includegraphics[width=8cm]{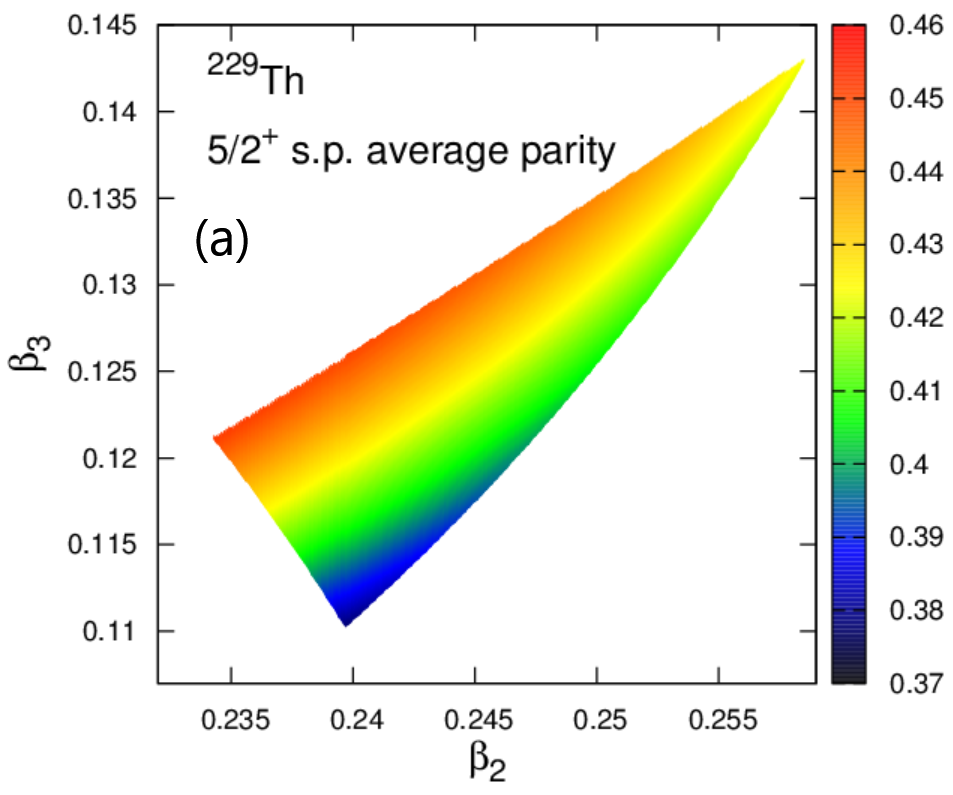}
\includegraphics[width=8cm]{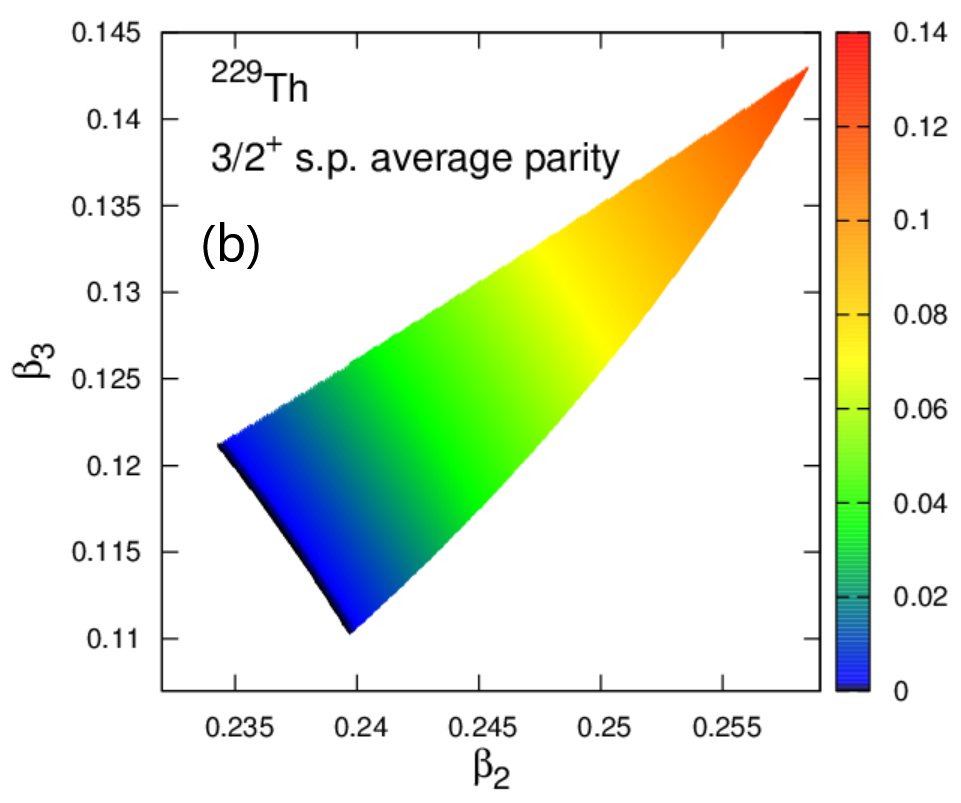}
\caption{Average parity $\langle\pi_{\mbox{\scriptsize sp}}\rangle$ in the (a)
GS and (b) IS s.p. orbitals appearing in DSM within the model-defined QO
deformation space.}
\label{th229_avpar_mod}
\end{figure*}

\begin{figure*}
\centering
\includegraphics[width=8cm]{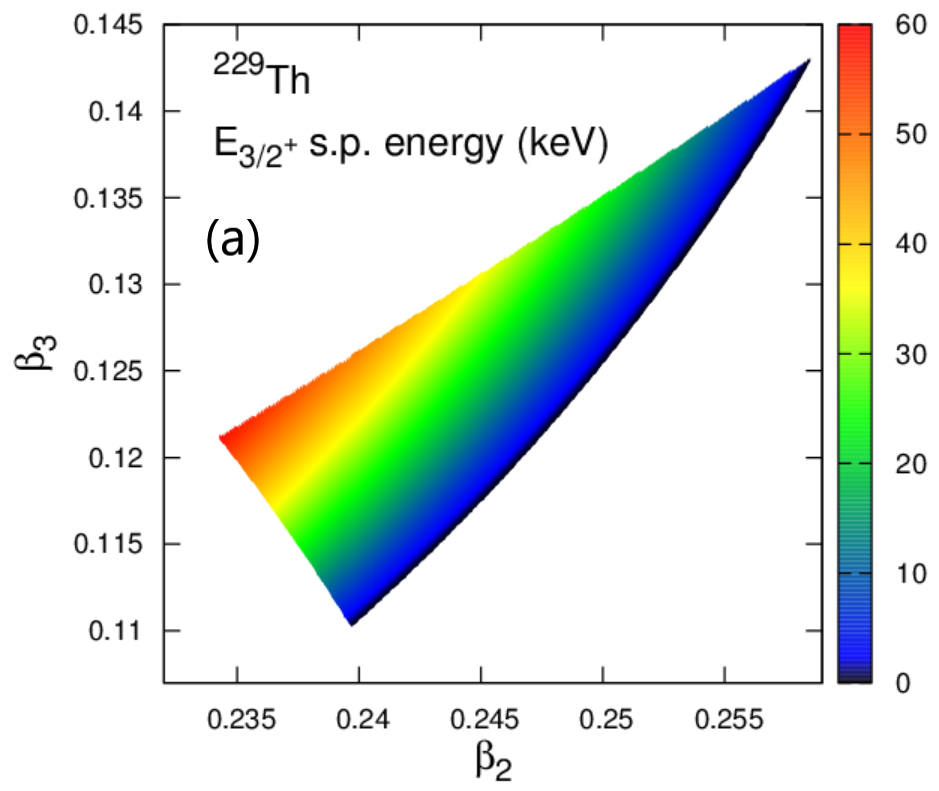}
\includegraphics[width=8cm]{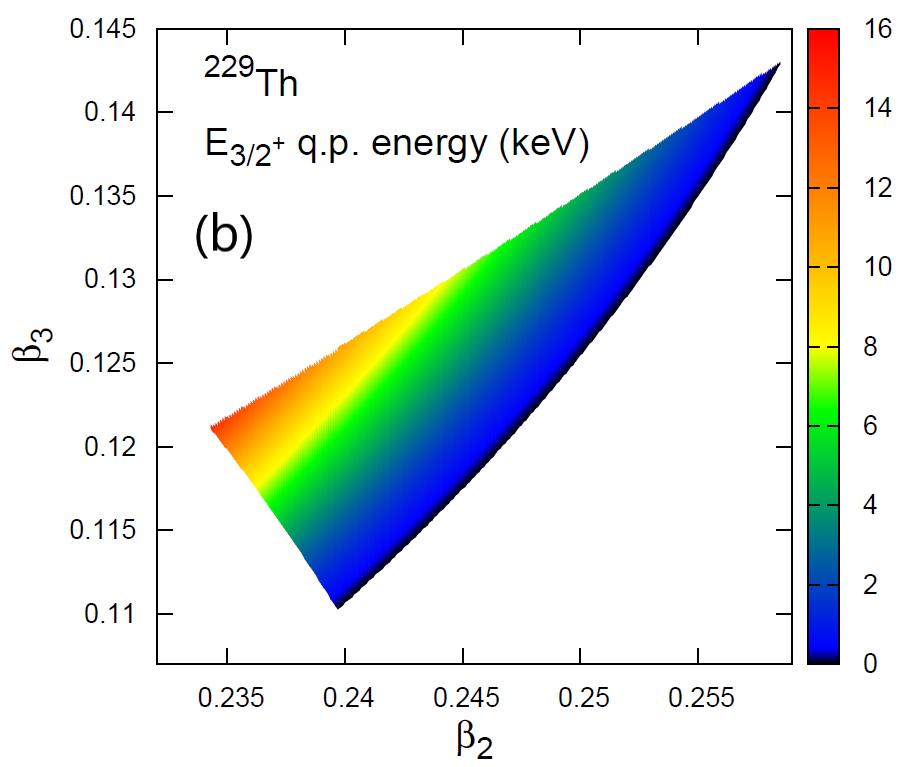}
\caption{(a) S.p. and (b) q.p. energy (in keV) of the $3/2^{+}$ isomer orbital
with respect to the $5/2^{+}$ GS orbital appearing in DSM within the
model-defined space of QO deformations. The q.p. energy is obtained with pairing
parameters $g_0=18.805$ MeV and $g_1=7.389$ MeV used in Eq.~(\ref{Gnp}). See
text for further explanations.}
\label{th229_spqpen}
\end{figure*}

\section{Numerical results and discussion}
\label{results}
\subsection{Determination of the deformed shell model deformation space}

The Woods-Saxon DSM shape parameters $\beta_2$ and $\beta_3$ represent a
basic input of our model and their values are decisive for the model
predictions. Hereafter under ``QO deformations and/or parameters'' we will
understand these two quantities unless otherwise specified. In
Ref.~\cite{Minkov_Palffy_PRL_2017} the quadrupole-deformation parameter
$\beta_2$ was chosen by varying it between the experimental values 0.230 and
0.244 available for the neighbouring even-even nuclei $^{228}$Th and
$^{230}$Th, respectively \cite{Raman2001}. Simultaneously the
octupole-deformation parameter $\beta_3$ was varied to obtain the GS and IS
orbitals very close to each other, with leading 5/2[633] and 3/2[631]
components in the respective s.p. wave-function expansions given in
Eq.~(\ref{ahodecomp}), and with positive average values of the parity
$\langle\pi_{\mbox{\scriptsize sp}}\rangle >0$ from Eq.~(\ref{avepar}) in
both s.p. states. We note that the chosen interval for the octupole
deformation was at that time solely relying on model estimates. However, in
the meantime we have found out that this range is also supported by an
independent microscopic result.  In Ref.~\cite{Nomura2014} self-consistent
relativistic Hartree-Bogoliubov model calculations with the universal energy
density functional DD-PC1 \cite{Nik2008} predict a rather deep total-energy
minima for $\beta_3$ between 0.1 and 0.2 in the neighboring even-even nuclei
$^{228}$Th and $^{230}$Th.

In this study we identify the ($\beta_2,\beta_3$) deformation space which
could provide a relevant model description of the $^{229m}$Th isomer similar
to the one obtained in
Refs.~\cite{Minkov_Palffy_PRL_2017,Minkov_Palffy_PRL_2019} which had
considered the values $\beta_2=0.240$ and $\beta_3=0.115$. To this end we
have performed DSM calculations on a grid in the ranges $0.2\leq\beta_{2}\leq
0.26$ and $0\leq\beta_{3}\leq 0.15$ which are supposed to include the QO
deformations physically relevant for a nucleus in the mass region of
$^{229}$Th. At each point of the grid we obtain the $K$-value and the average
parity $\langle\pi_{\mbox{\scriptsize sp}}\rangle$ for the last occupied s.p.
orbital, which is supposed to determine the GS and for the next (first)
non-occupied orbital, candidate for the IS. Note that the calculation does
not involve the collective (CQOM) part of the model and the only entering
parameters are the two Woods-Saxon DSM deformations.

\begin{figure*}
\centering
\includegraphics[width=8cm]{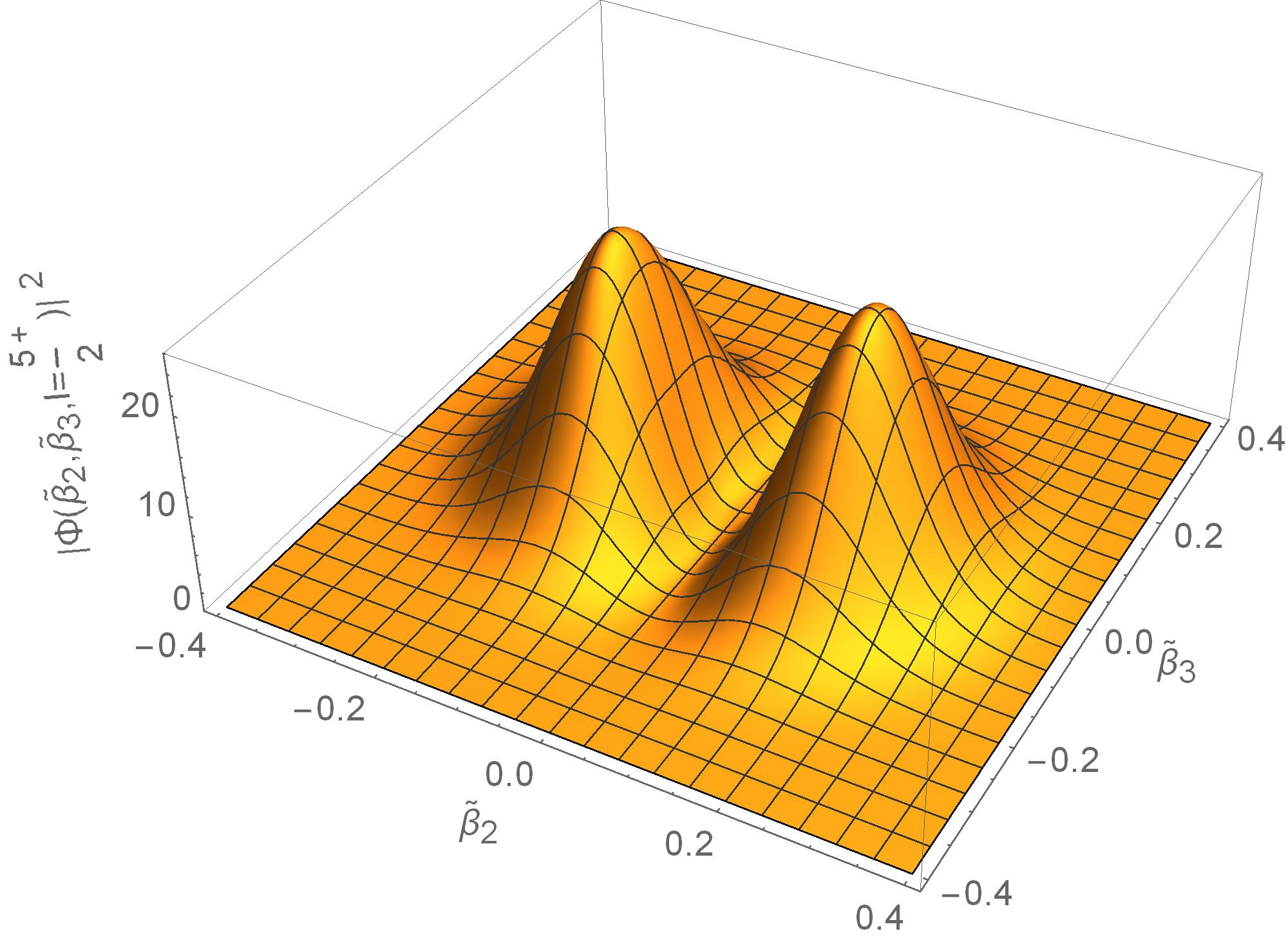}
\includegraphics[width=8cm]{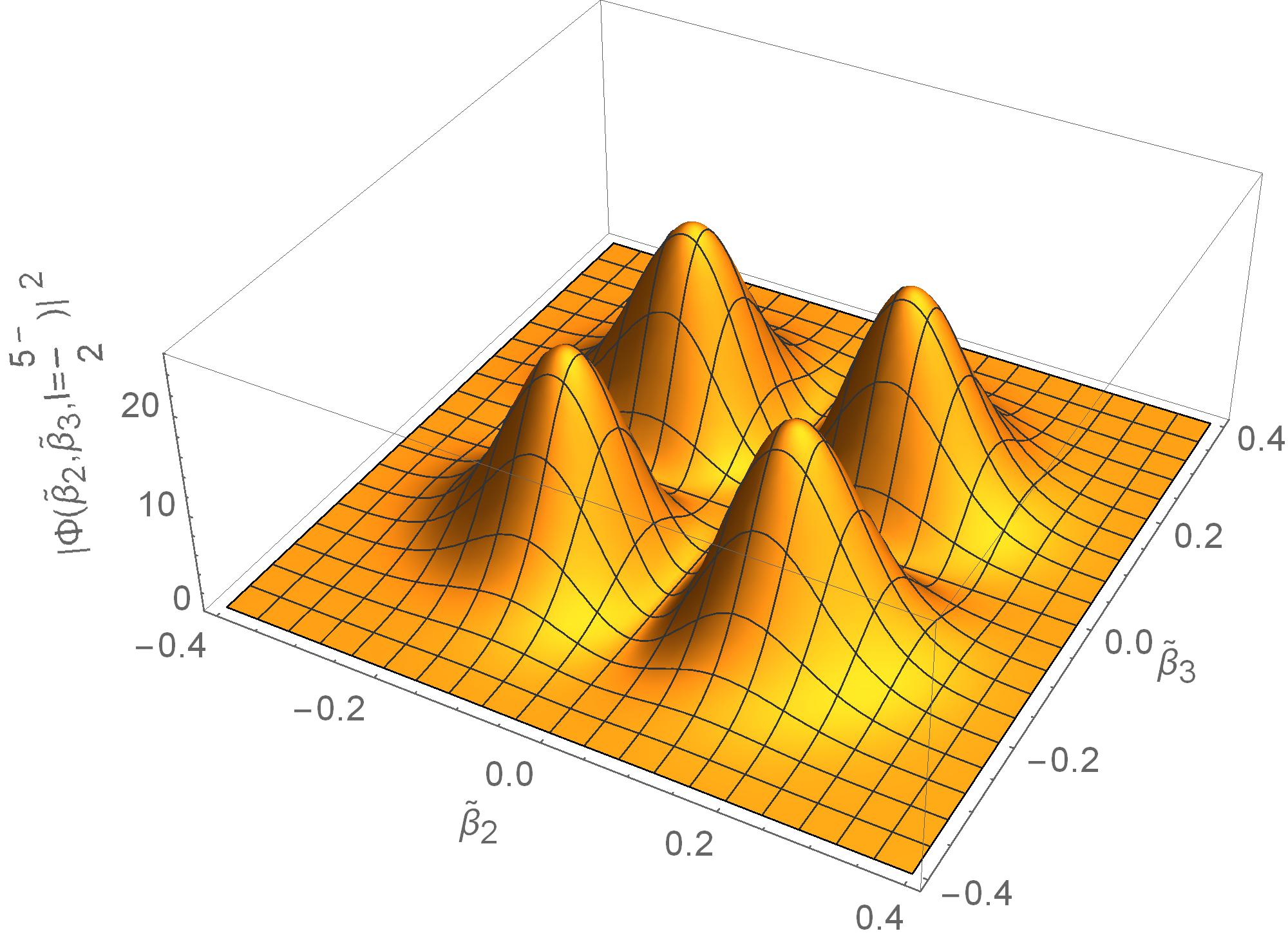}
\includegraphics[width=8cm]{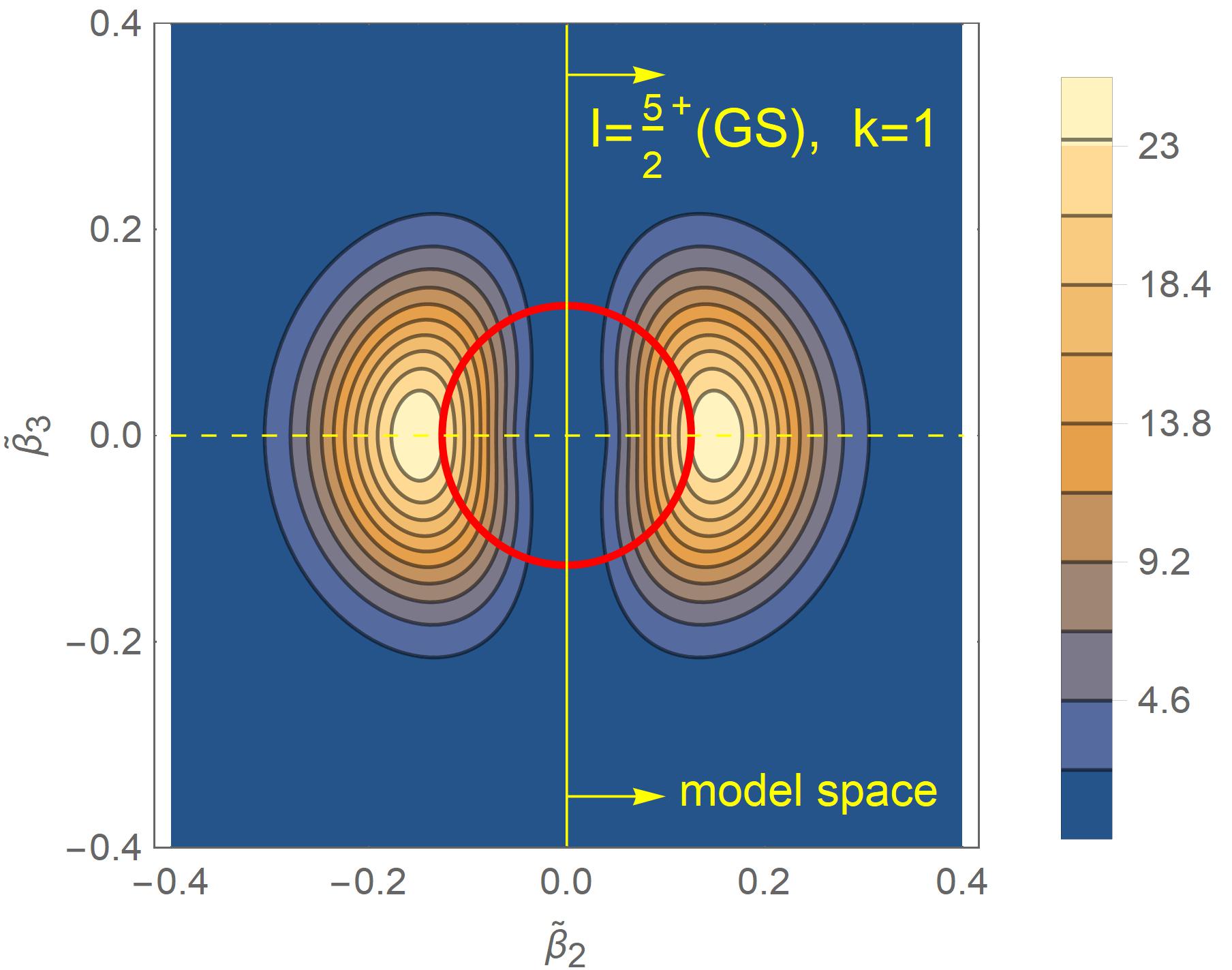}
\includegraphics[width=8cm]{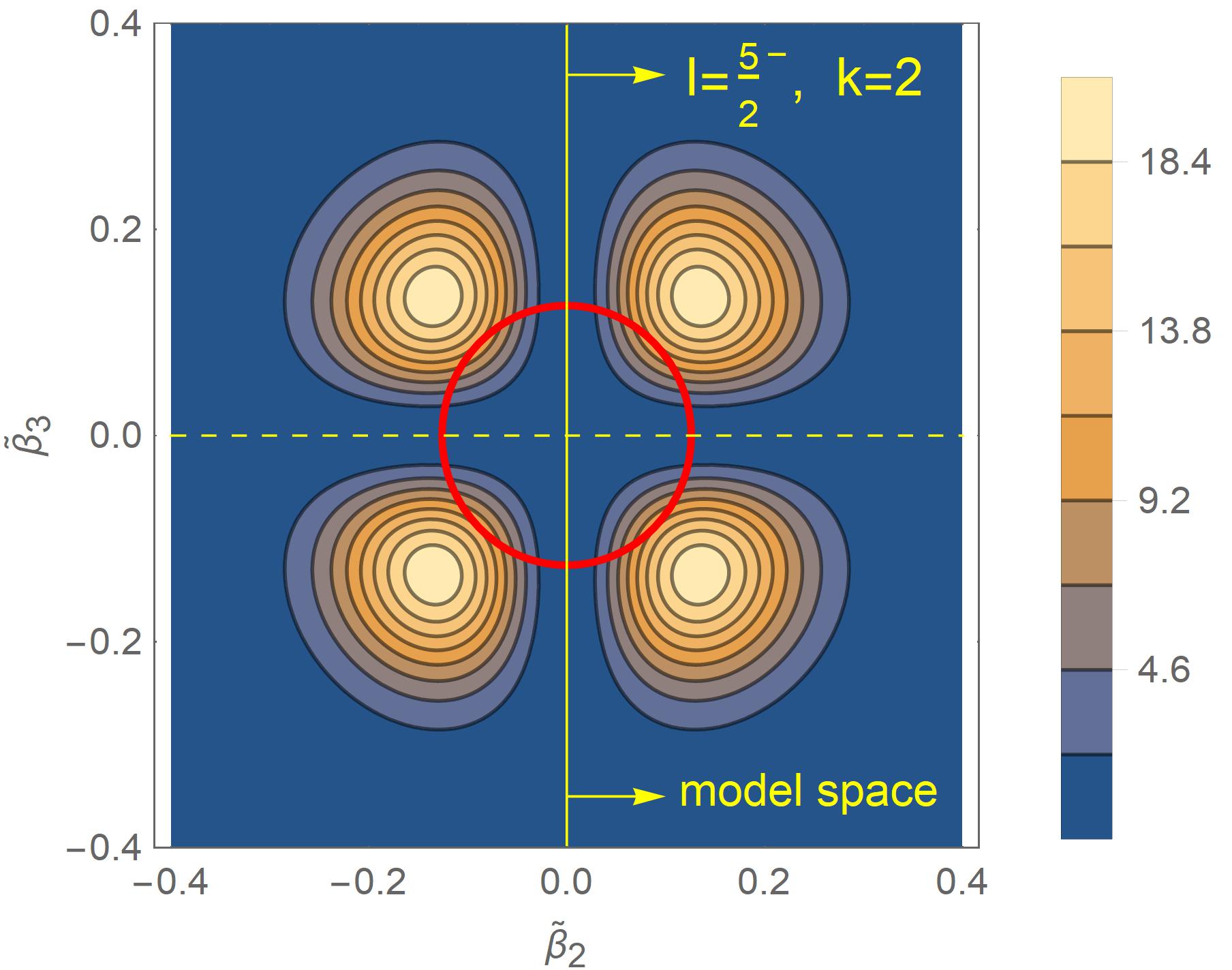}
\caption{The CQOM QO wave-function density
$|\Phi^{\pi_{\mbox{\scriptsize qo}}}_{nkI}(\tilde{\beta_2},\tilde{\beta_3})|^{2}$
from Eq.~(\ref{wvib}) as a function of $\tilde{\beta}_{2}$ and $\tilde{\beta}_{3}$
for the $I^{\pi}=5/2^{+}$ GS (with $k=1$) and $5/2^{-}$ state (with $k=2$) of the
yrast quasi-parity-doublet in $^{229}$Th. We use here the DSM+CQOM fit with
$q_s=q_R=0.6$ and the Woods-Saxon DSM deformation parameters
($\beta_2,\beta_3$)=($0.240,0.115$). The upper panels represent three-dimensional
plots while the lower panels illustrate the corresponding projected two-dimensional
contour plots. The CQOM potential bottoms defined by the semiaxes
$\tilde{\beta}_{2,3}^{\mbox{\scriptsize sa}}$ are shown as red circles. }
\label{wfdensities}
\end{figure*}

\begin{figure*}
\centering
\includegraphics[width=8cm]{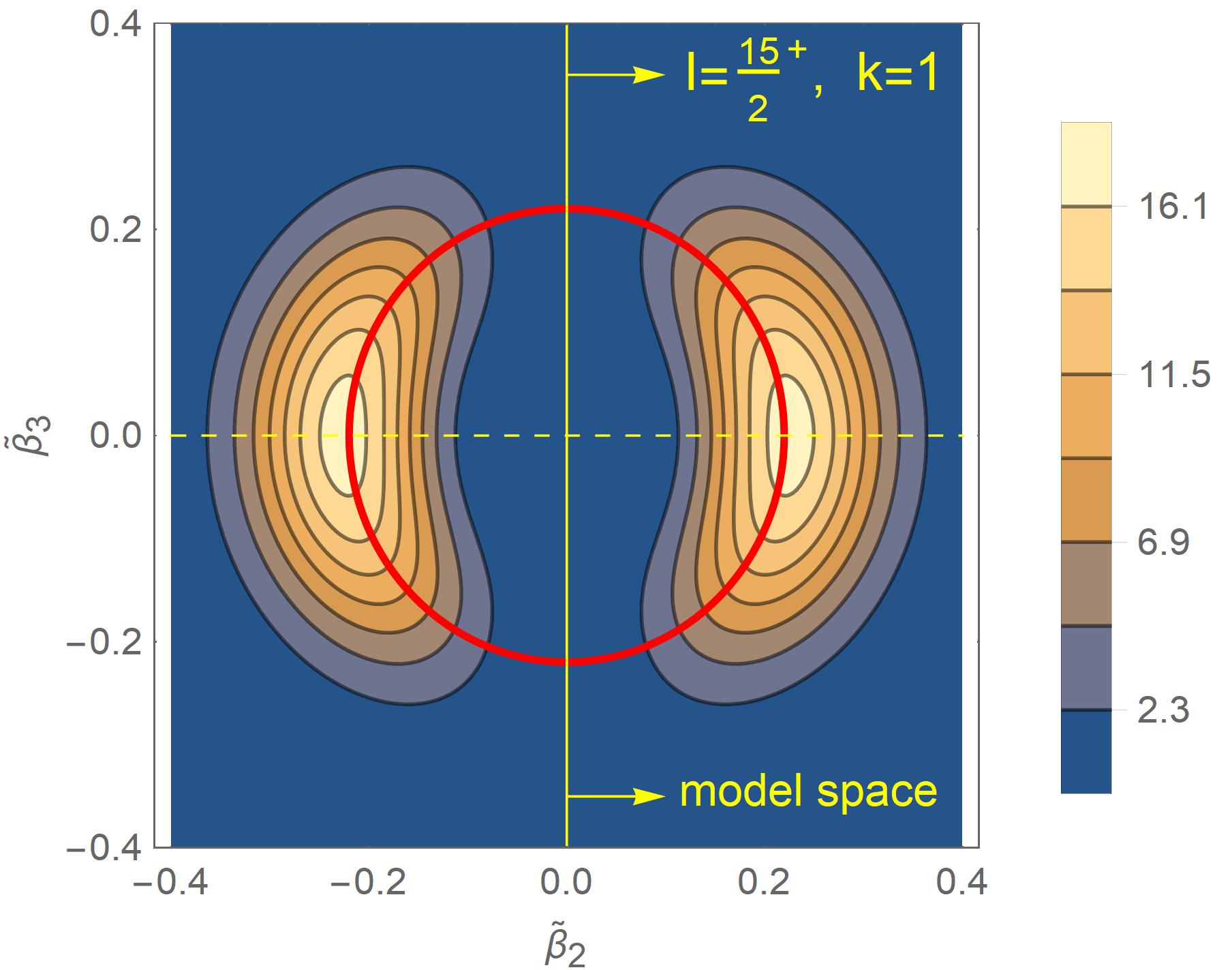}
\includegraphics[width=8cm]{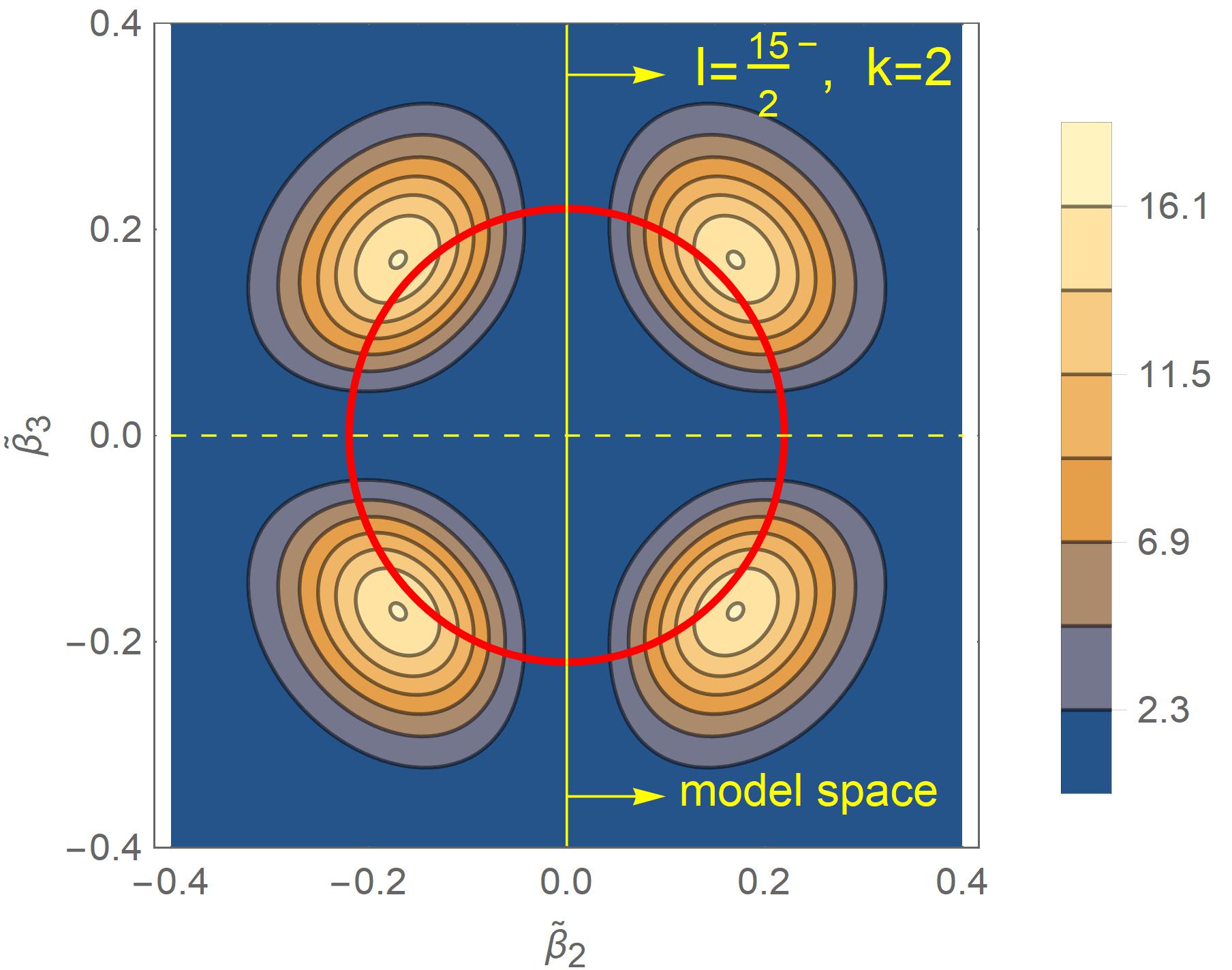}
\caption{The same as Fig.~\ref{wfdensities} (lower panels), but for the
$I^{\pi}=15/2^{+}$ and $15/2^{-}$ yrast quasi-parity-doublet states of
$^{229}$Th. Note the considerable increase in the QO semiaxes and the
corresponding expansion of the dynamical-deformation peak positions in
the space of the collective CQOM variables $\tilde{\beta}_2$ and
$\tilde{\beta}_3$.}
\label{wfdenscont15div2}
\end{figure*}

The result of this calculation is shown in Fig.~\ref{th229_kpi}.
Figs.~\ref{th229_kpi} (a) and \ref{th229_kpi} (b) present in color coding the
$(\beta_2,\beta_3)$ areas in which different $K$ values appear for the GS and
IS orbitals, respectively. For the GS orbital the $K=5/2$ value appears in
two (yellow) regions, while for the IS the $K=3/2$ value appears in one
narrow (blue) region. The intersection of the $K=5/2$ GS and $K=3/2$ IS
subspaces coincides with the blue $K=3/2$ region for the IS and depicts the
($\beta_2,\beta_3$)- region in which the DSM provides the required 5/2[633]
and 3/2[631] orbitals for the GS and IS, respectively. Furthermore,
considering the information from Fig.~\ref{th229_kpi} (c) and \ref{th229_kpi}
(d), one can identify in a similar way the regions with positive and negative
average values of the parity in the GS and IS orbitals, respectively. By
retaining only the $\langle\pi_{\mbox{\scriptsize sp}}\rangle >0$ areas for
both orbitals, one ends up with a rather limited ($\beta_2,\beta_3$)-region
given by the thick triangle contour in the four plots. This region includes
all QO deformations from the considered space which are relevant within the
DSM regarding the current experimental information and theoretical
interpretation of the $K^{\pi}=3/2^{+}$ isomer in $^{229}$Th. Hereinafter we
call this region our ``model deformation space''.

Based on the above result, we can draw the following important conclusion.
Considering the long-adopted $K$-value and parity of the $^{229m}$Th IS, our
DSM prediction shows that this isomer can only exist at essentially nonzero
octupole deformation of the s.p. potential. More precisely, one can say that
the coexistence of the $K^{\pi}=3/2^{+}$ IS together with the
$K^{\pi}=5/2^{+}$ GS requires the presence of nonzero octupole deformation,
as seen from Fig.~\ref{th229_kpi}(a). In fact our more extended calculations
in the QO deformation grid show that for $\beta_2 < 0.2$ the (yellow) range
of coexisting $K^{\pi}=5/2^{+}$ GS and $K^{\pi}=3/2^{+}$ IS orbitals goes
down and further reaches the $\beta_3 = 0$ line. However, this occurs around
$\beta_2 \sim 0.1$, which is far beyond the deformation limits typical for
this mass region. Thus, we can conclude that the octupole deformation appears
to be of a crucial importance for the formation of the $^{229m}$Th isomer
according to the present knowledge on the corresponding GS and IS angular
momenta and parities.

Furthermore, we note that the deformation values
($\beta_2=0.240,\beta_3=0.115$) used in
Refs.~\cite{Minkov_Palffy_PRL_2017,Minkov_Palffy_PRL_2019} appear close to
the lowest vertex of the investigated DSM deformation space. The relatively
small area of this space suggests a reasonable degree of arbitrariness in the
model conditions imposed in the studies of
Refs.~\cite{Minkov_Palffy_PRL_2017} and \cite{Minkov_Palffy_PRL_2019}
regarding the choice of QO deformation. Moreover, the deformation region
determined in this work  appears to be consistent with the corresponding
areas of the QO minima in the energy surfaces of $^{228}$Th and $^{230}$Th
obtained in the relativistic Hartree-Bogoliubov  model calculations
\cite{Nomura2014}. Nevertheless, the precise determination and prediction of
the $^{229m}$Th isomer properties as well as the deeper understanding of the
mechanism governing its formation requires a more detailed examination of the
model descriptions obtained for various deformations fixed in the outlined
model space. In the following our study is focused on this task.

A direct consequence of the location of the model space at nonzero octupole
deformation is that the GS and IS s.p. orbitals provided by the DSM always
appear with mixed parity which has to be projected in the total core plus
particle wave function, as seen in Eq.~(\ref{wfpcore}), and implemented in
the model procedure applied in \cite{Minkov_Palffy_PRL_2017}. The average
parity $\langle\pi_{\mbox{\scriptsize sp}}\rangle$ of both orbitals
calculated from Eq.~(\ref{avepar}) as a function of the QO deformations
within the model space is illustrated in Fig.~\ref{th229_avpar_mod}. The
results show that while in the GS the quantity $\langle\pi_{\mbox{\scriptsize
sp}}\rangle$ varies within the limits $0.37-0.46$, in the IS the parity
mixing is even much stronger with $\langle\pi_{\mbox{\scriptsize sp}}\rangle$
varying between 0 and 0.14. The black side of the triangle in
Fig.~\ref{th229_avpar_mod}(b) corresponds to the (left) border of the space
where the average parity of the IS turns from positive to negative values.
This result shows that the mechanism governing the formation of the isomer is
even more complicated due to the fine parity-mixed structure of the s.p. wave
functions and the accordingly applied projection procedure.

\begin{figure}
\centering
\includegraphics[width=8cm]{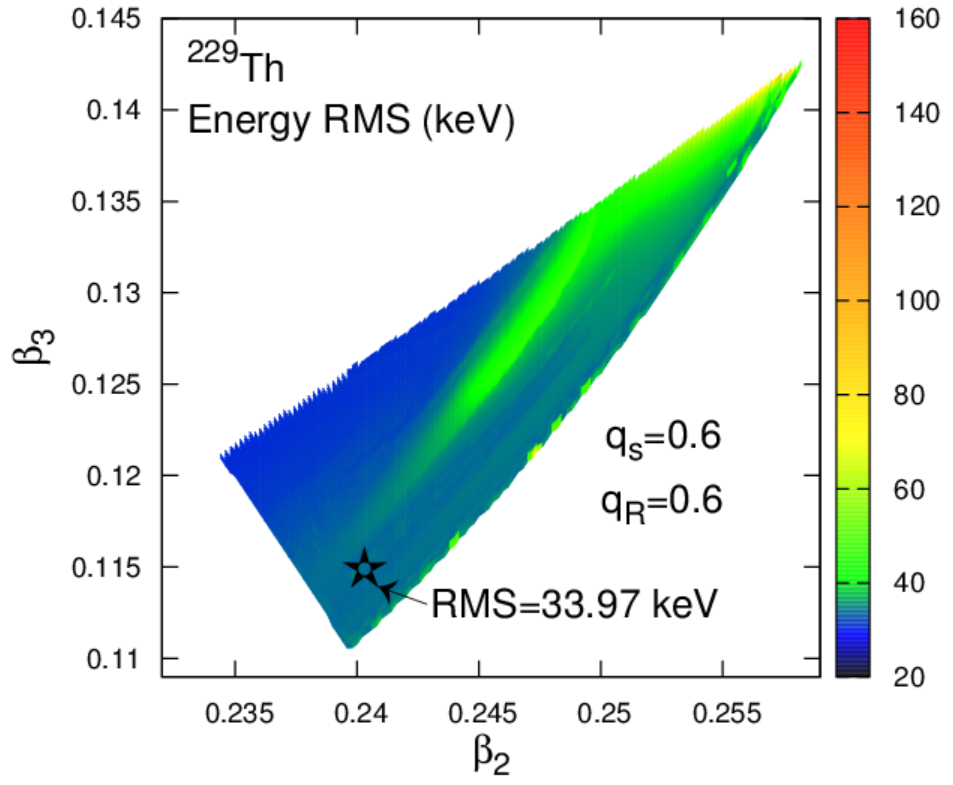}
\caption{Energy RMS values in keV for the GS (yrast) and IS (excited) bands together
obtained by the model fit on a grid within the model-defined QO deformation space.
We use $q_s=q_R=0.6$. The open star indicates the location of the deformations
($\beta_2,\beta_3$)=($0.240,0.115$) adopted in
Refs.~\cite{Minkov_Palffy_PRL_2017,Minkov_Palffy_PRL_2019}.}
\label{th229_rms}
\end{figure}

\begin{figure*}
\centering
\includegraphics[width=8cm]{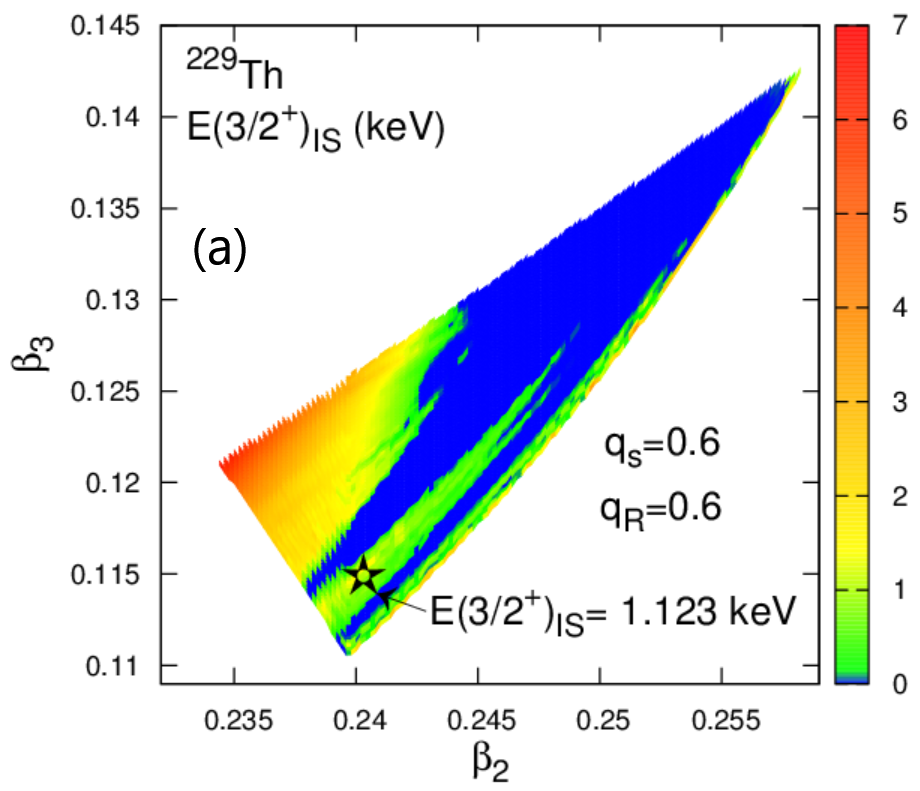}
\includegraphics[width=8cm]{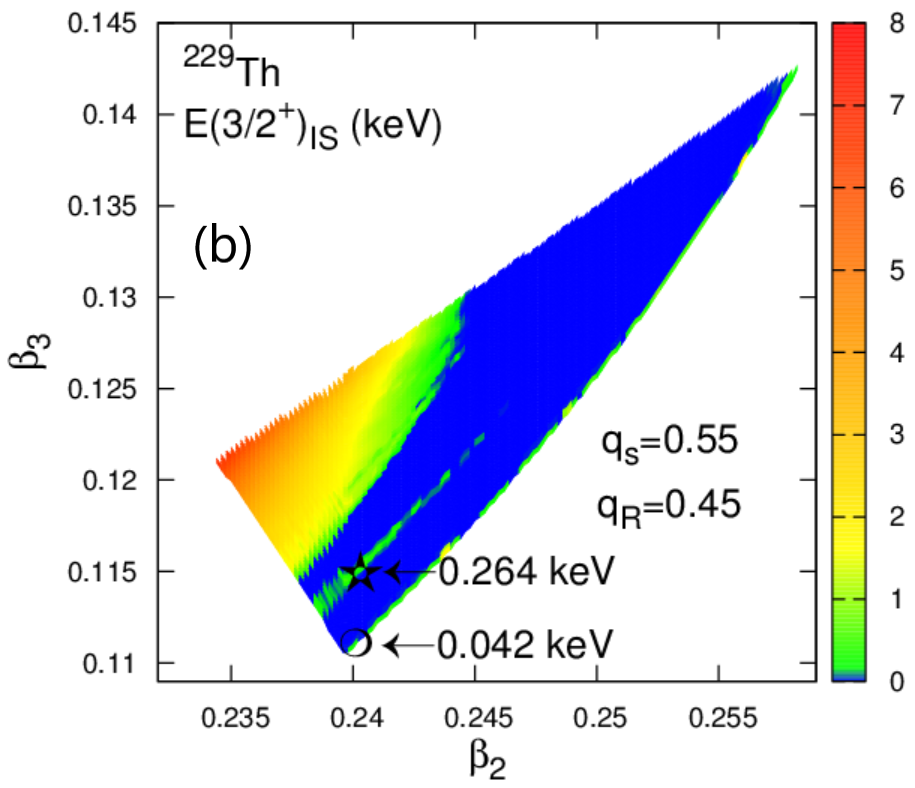}
\caption{Isomer energy $E_{IS}$ obtained by the model fits on a model-defined
QO deformation space grid for two different combinations of $q_s$ and $q_R$.
The open star indicates the location of the deformations
($\beta_2,\beta_3$)=($0.240,0.115$) adopted in
Refs.~\cite{Minkov_Palffy_PRL_2017,Minkov_Palffy_PRL_2019}.
The circle indicates the set ($\beta_2,\beta_3$)=($0.240,0.111$) situated closer
to the degeneracy line.}
\label{th229_eniso}
\end{figure*}

Other important quantities delivered by the DSM are the s.p. and q.p.
energies for the IS orbital determined with respect to the corresponding
energies in the GS orbital, $E^{\mbox{\scriptsize sp}}_{3/2^{+}}=
E_{\mbox{\scriptsize sp}}^{3/2^{+}}-E_{\mbox{\scriptsize sp}}^{5/2^{+}}$ and
$E^{\mbox{\scriptsize qp}}_{3/2^{+}}= \epsilon_{\mbox{\scriptsize
qp}}^{3/2^{+}}-\epsilon_{\mbox{\scriptsize qp}}^{5/2^{+}}$. The lowering of
the q.p. energy with respect to the s.p. energy can be controlled through
additional tuning of the pairing constants as shown in
Ref.~\cite{Minkov_Palffy_PRL_2017}. However, as argued at the end of
Sec.~\ref{modappl}, here we use the $g_0$ and $g_1$ values fixed in
\cite{Minkov_Palffy_PRL_2017} focusing our analysis on the deformation
dependencies. In Fig.~\ref{th229_spqpen} both $E^{\mbox{\scriptsize
sp}}_{3/2^{+}}$ and $E^{\mbox{\scriptsize qp}}_{3/2^{+}}$ for the IS are
plotted as functions of $\beta_2$ and $\beta_3$. As expected, they show an
identical dependence but with different nominal values. In addition, along
the right side of the triangle the q.p. and s.p. content of the isomer energy
goes to zero, i.e. the two orbitals, 5/2[633] and 3/2[631] mutually
degenerate. This is an important limit of the model deformation space. In
fact the black lines in Fig.~\ref{th229_spqpen} correspond to the crossing of
both orbitals when leaving the model space to enter the blue area with
$K=3/2$ in Fig.~\ref{th229_kpi}(a) and the lower $K=5/2$ (yellow) area in
Fig.~\ref{th229_kpi}(b), a situation in which the GS appears with
$K^{\pi}=3/2^{+}$ and the IS obtains $K^{\pi}=5/2^{+}$. The proximity to this
line from the model space interior determines the degree of the q.p.
quasidegeneracy effect. For the pair of QO deformations
($\beta_2,\beta_3$)=($0.240,0.115$) adopted in
Refs.~\cite{Minkov_Palffy_PRL_2017,Minkov_Palffy_PRL_2019}, the $3/2^{+}$
q.p. energy yields $E^{\mbox{\scriptsize qp}}_{3/2^{+}}=2.196$ keV. We note
that this is not the final IS energy in which additional contributions take a
part as explained in Sec.~\ref{modappl}. The upper side of the triangle
corresponds to the crossing of the $K^{\pi}=3/2^{+}$ orbital with a
$K^{\pi}=7/2^{-}$ orbital with leading component 7/2[743] [see red area in
Fig.~\ref{th229_kpi}(b)]. It is not of a particular interest from the
isomer-formation point of view.

\subsection{Coherent quadrupole-octupole model fits in the deformed shell
model deformation space}
\label{cqomfits}

At this point we are ready to examine the behaviour of the model description
and prediction for the physical observables of interest within the DSM
deformation space. We are especially interested in the corresponding
behaviour of  the $B(M1)$ and $B(E2)$ IS transition rates and of the IS and
GS magnetic moments $\mu_{\mbox{\scriptsize GS}}$ and $\mu_{\mbox{\scriptsize
IS}}$. To this end we have performed full model fits by adjusting the five
CQOM parameters, $\omega$, $b$, $d_0$, $c$, $p$ and the Coriolis mixing
constant $A$ with respect to the experimental quasi-parity-doublet spectrum,
the available transition rates and magnetic moments at each point of the
deformation space grid with the pairing constants fixed as described above.
Thus, for each pair of Woods-Saxon DSM QO deformations we obtain the full
spectroscopic description of the nucleus storing the quantities of interest
for our systematic analysis presented below.

\begin{figure*}[ht]
\centering
\includegraphics[width=8cm]{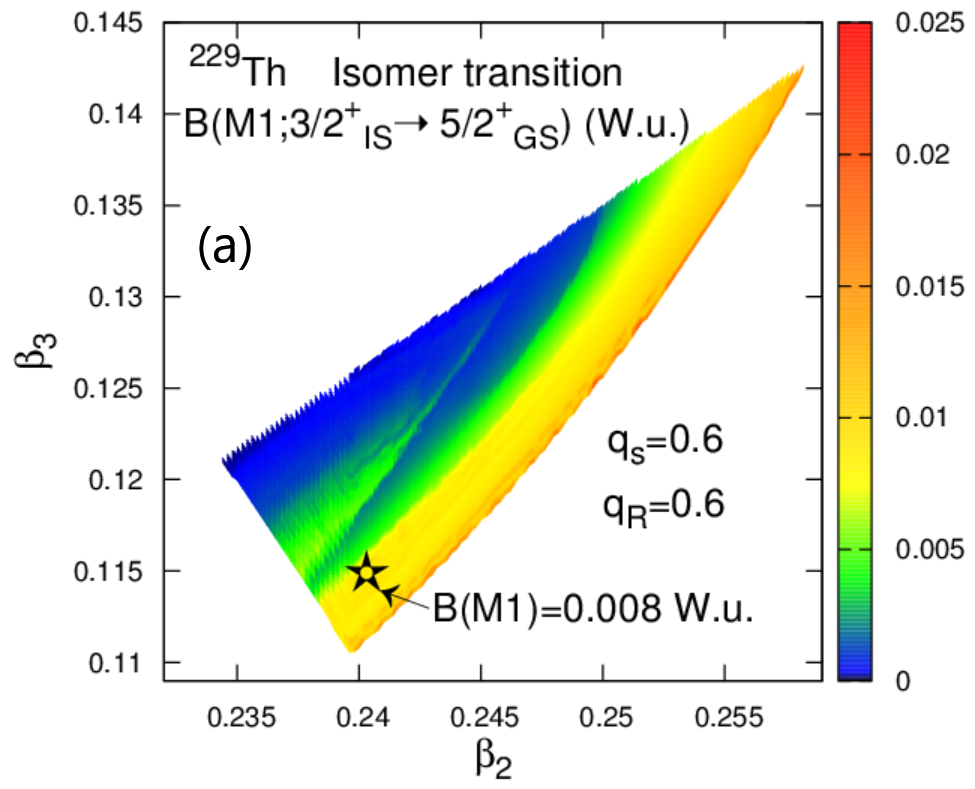}
\includegraphics[width=8cm]{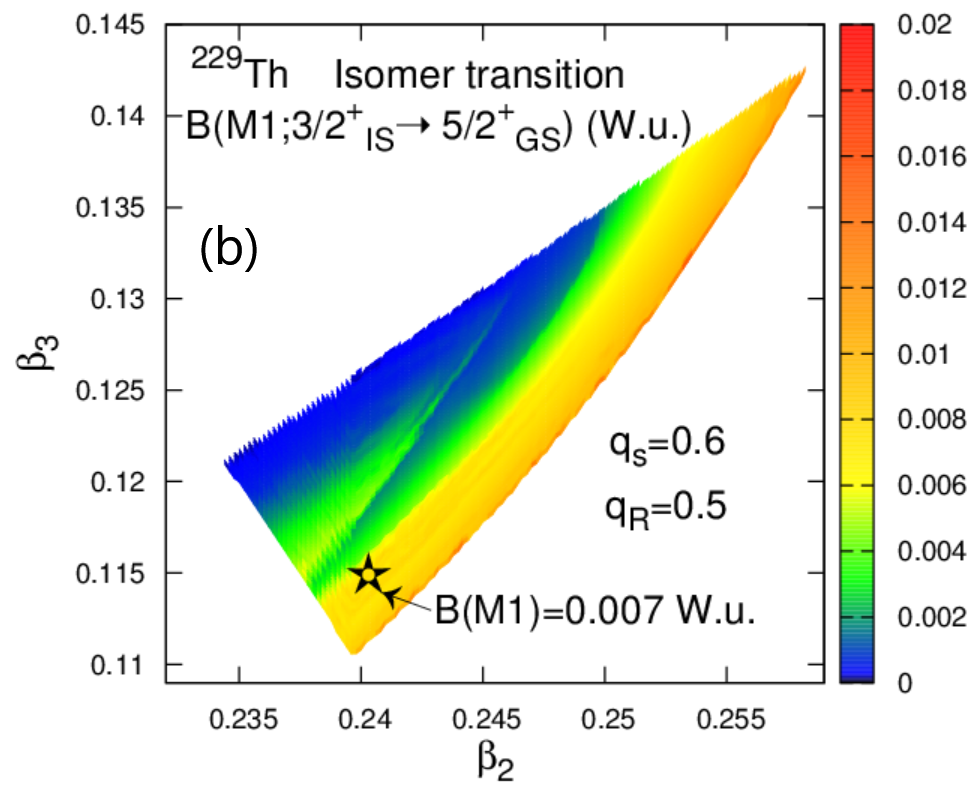}
\includegraphics[width=8cm]{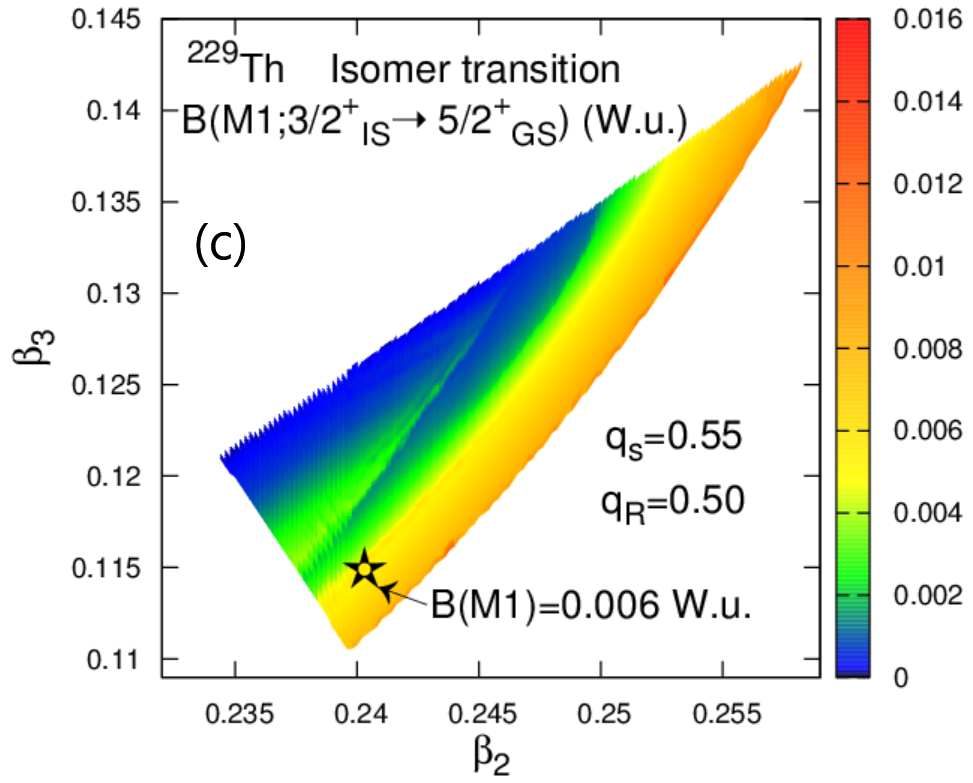}
\includegraphics[width=8cm]{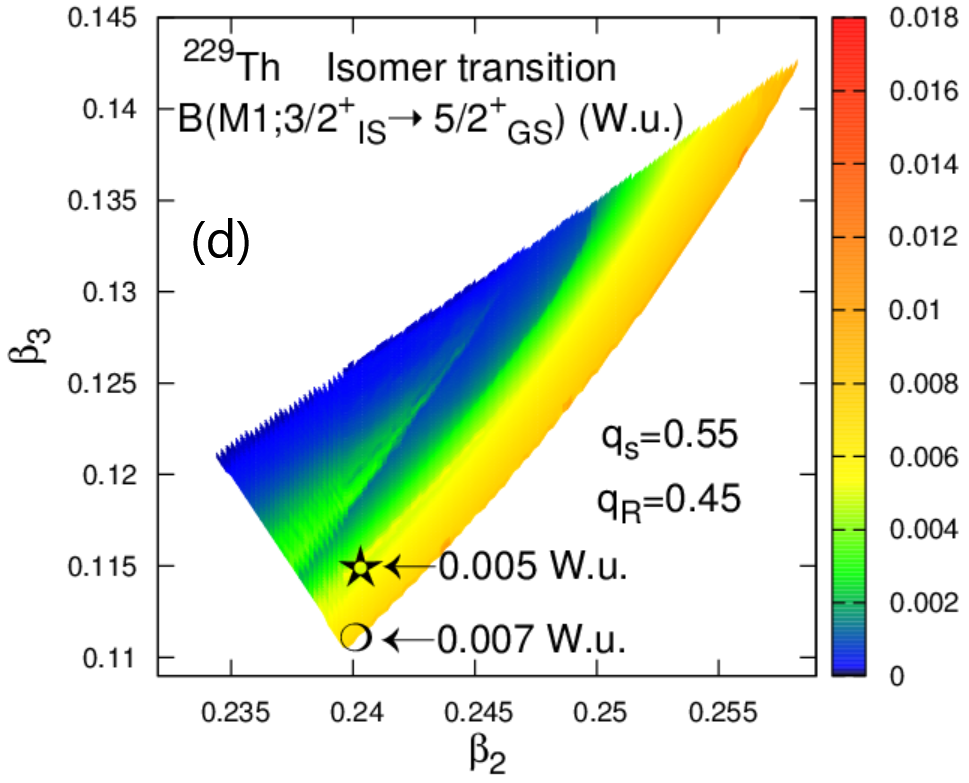}
\caption{$B(M1)$ isomer transition values obtained by the model fits on a grid
within the model-defined QO deformation space at four different combinations of
$q_s$ and $q_R$. The open star  indicates the location of the deformations
($\beta_2,\beta_3$)=($0.240,0.115$) adopted in
Refs.~\cite{Minkov_Palffy_PRL_2017,Minkov_Palffy_PRL_2019}. The circle in panel
(d) indicates the set ($\beta_2,\beta_3$)=($0.240,0.111$) situated closer to
the degeneracy line.}
\label{th229_m1isotrans}
\end{figure*}

\begin{figure*}[ht]
\centering
\includegraphics[width=8cm]{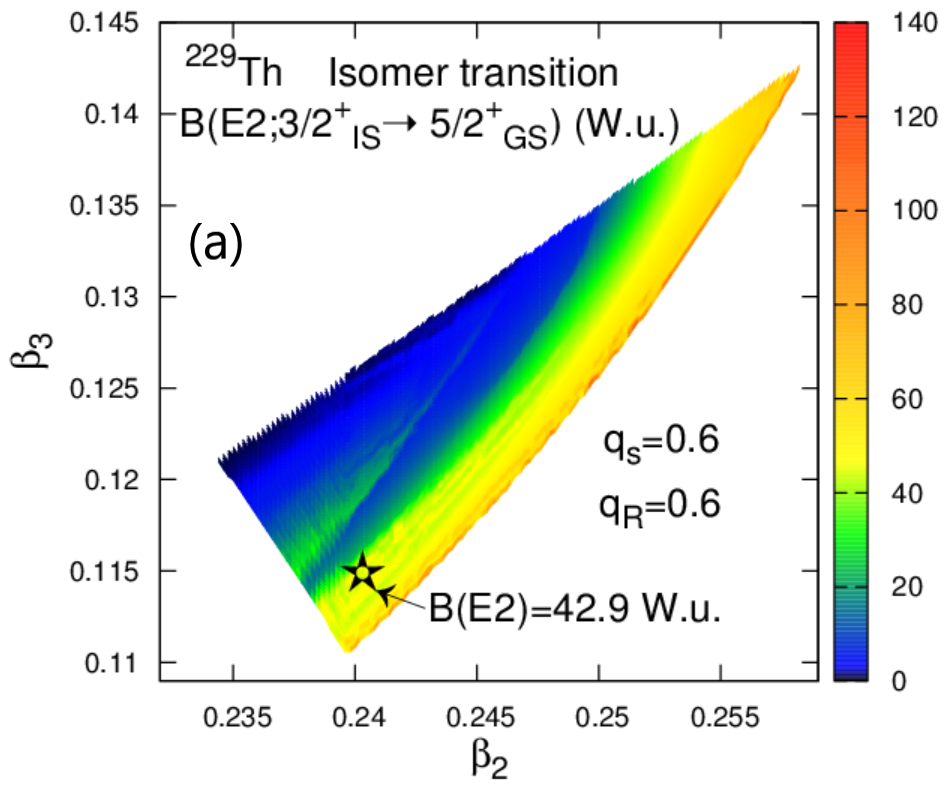}
\includegraphics[width=8cm]{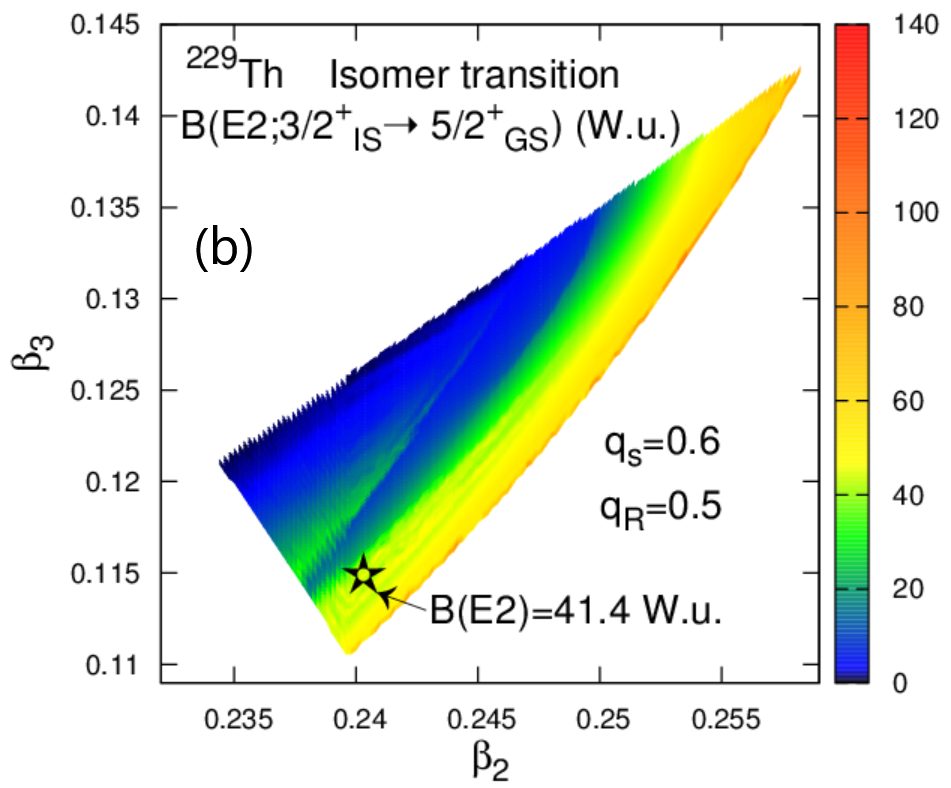}
\includegraphics[width=8cm]{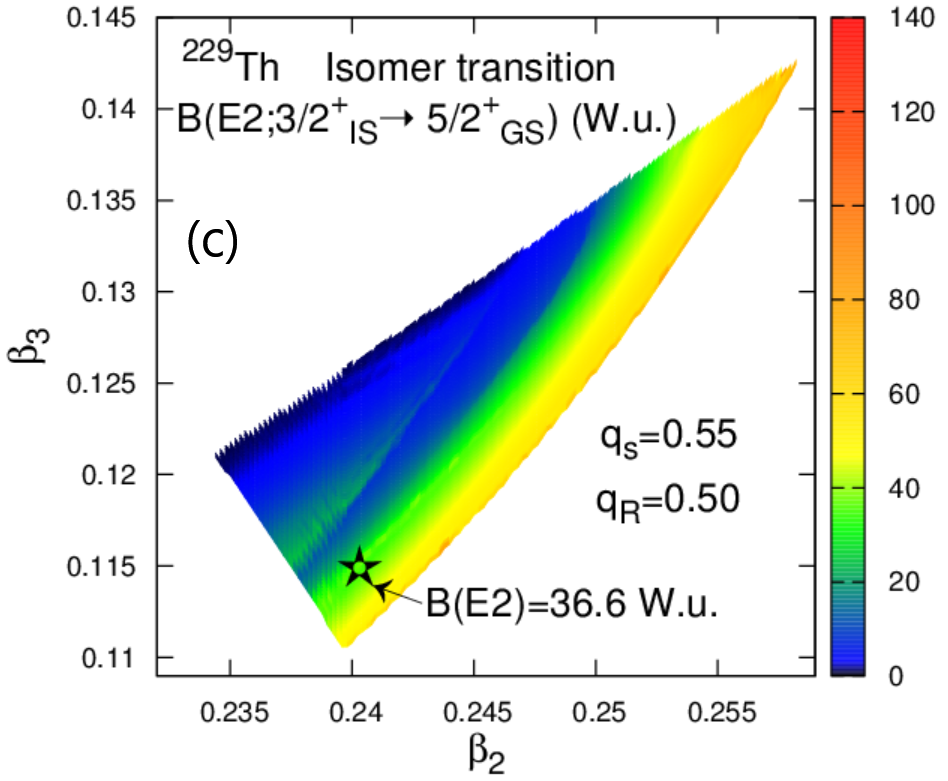}
\includegraphics[width=8cm]{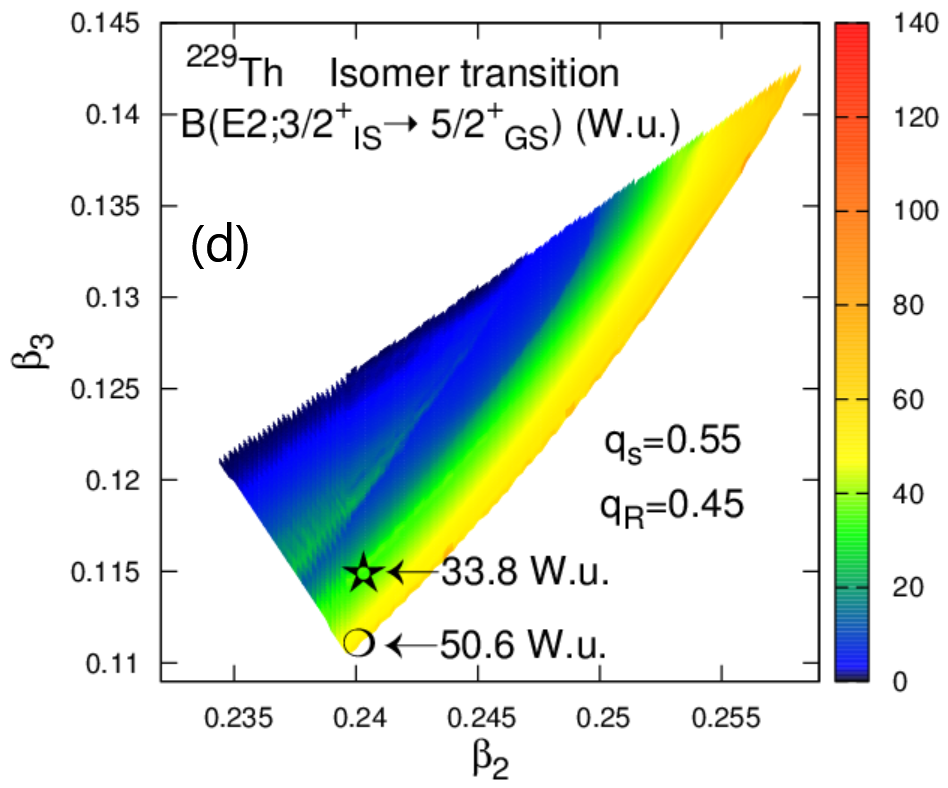}
\caption{$B(E2)$ isomer transition values obtained by the model fits
on the model-defined QO deformation space grid at four different combinations
of $q_s$ and $q_R$. The open star  indicates the location of the
deformations ($\beta_2,\beta_3$)=($0.240,0.115$) adopted in
Refs.~\cite{Minkov_Palffy_PRL_2017,Minkov_Palffy_PRL_2019}. The circle in panel
(d) indicates the set ($\beta_2,\beta_3$)=($0.240,0.111$) situated closer to
the degeneracy line.}
\label{th229_e2isotrans}
\end{figure*}

The five adjusted CQOM parameters show a smooth behavior along the
deformation space with values consistent with those obtained in
Ref.~\cite{Minkov_Palffy_PRL_2017}. We therefore refrain from addressing
further numerical details  here.  We note however that the parameter $p$
defined in Eq.~(\ref{pqd}) is close to unity $p\approx 1$ throughout the
entire DSM ($\beta_2,\beta_3$)-model space. This parameter indicates the
relative contribution of the quadrupole and octupole modes in the CQOM. Thus,
according to Eq.~(\ref{semrat}) all parameter fits lead to a practically
circular bottom of the CQOM potential with
$\tilde{\beta}_{2}^{\mbox{\scriptsize
sa}}\approx\tilde{\beta}_{3}^{\mbox{\scriptsize sa}}$, showing that the model
describes the collective quasi-parity-doublet structure of the $^{229}$Th
spectrum with equal weights of the quadrupole $\tilde{\beta}_{2}$ and
octupole $\tilde{\beta}_{3}$ deformation modes. Considering the DSM
deformation parameters  $\beta_2=0.240$ and $\beta_3=0.115$ used in
Ref.~\cite{Minkov_Palffy_PRL_2017}, we obtain for the $5/2^{+}$ GS and
$3/2^{+}$ IS states values between 0.12 and 0.17 for the two equal
$\tilde{\beta}_{2,3}^{\mbox{\scriptsize sa}}$  semiaxes (which define the
circle radius of the CQOM potential bottom).

Inspection of the odd-nucleon contribution to
$\tilde{\beta}_{2,3}^{\mbox{\scriptsize sa}}$ determined in
Eq.~(\ref{semiaxtilde}) through Eq.~(\ref{Xmix}) at
($\beta_2,\beta_3$)=($0.240,0.115$) shows that in the  case of $^{229}$Th the
Coriolis mixing causes a negligible decrease in
$\tilde{\beta}_{2,3}^{\mbox{\scriptsize sa}}$ compared with the core case,
Eq.~(\ref{semiax}), while a considerable decrease is caused by the term
$(-K_{b}^2)$. Thus while in the core+particle case the GS
$\tilde{\beta}_{2,3}^{\mbox{\scriptsize sa}}=0.122$, for the core only
[without the term $(-K_{b}^2)$] these values become 0.149. Similarly, in the
isomeric state the core+particle $\tilde{\beta}_{2,3}^{\mbox{\scriptsize
sa}}=0.121$, while in the core case the semiaxes values rise to 0.132.
Figure~\ref{wfdensities} illustrates the CQOM QO wave-function densities
$|\Phi^{\pi_{\mbox{\scriptsize
qo}}}_{nkI}(\tilde{\beta_2},\tilde{\beta_3})|^{2}$ from Eq.~(\ref{wvib}) for
the $I^{\pi}=5/2^{+}$ GS and its negative-parity counterpart $5/2^{-}$
obtained with the parameters of the CQOM fit at
($\beta_2,\beta_3$)=($0.240,0.115$). For simplicity we have only taken into
account the $(-K_{b}^2)$ term in $\widetilde{X}(I^{\pi},K_{b})$ dropping the
negligible Coriolis mixing term. The wave function density was calculated for
the quenching parameter set ($q_s,q_R$)=($0.6,0.6$). The dynamical
deformations are indicated by the positions of the density maxima in the
Figure. We see that in the two states these deformations appear outside of
the potential bottom circle. Furthermore, the dynamical deformation
parameters are not coinciding with the DSM ($\beta_2,\beta_3$) parameters,
confirming the discussion in Sec.~\ref{wf} on the distinction between
dynamical deformation parameters in the CQOM and the DSM deformation
parameters. We also note that the CQOM potential bottom semiaxes
$\tilde{\beta}_{2,3}^{\mbox{\scriptsize sa}}$ (the red circles) do not change
between the positive- and negative-parity counterparts in the
quasi-parity-doublet since the Coriolis mixing term only mixes states with
the same (bandhead) parity. This situation, however, would be different in a
spectrum build on the $K_{b}=1/2$ bandhead (which is not present for the case
of $^{229}$Th) where the decoupling term in $\widetilde{X}(I^{\pi},K_{b})$ in
Eq.~(\ref{Xmix}) would act in opposite directions on the semiaxes lengths of
the opposite-parity counterparts.

We have checked that the density plots for the $I^{\pi}=3/2^{+}$ IS and its
$3/2^{-}$ quasi-parity-doublet counterpart (not given here) look very similar
to those in Fig.~\ref{wfdensities}. To investigate the effect of higher
angular momentum values we show the CQOM QO wave-function densities for the
$I=15/2^{\pm}$ states of the yrast quasi-parity-doublet in
Fig.~\ref{wfdenscont15div2}. The CQOM potential semiaxes
$\tilde{\beta}_{2,3}^{\mbox{\scriptsize sa}}$ and the corresponding dynamical
deformations considerably increase with angular momentum, reaching at
$I=15/2^{\pm}$ values larger than $0.2$. This shows that the dynamical
deformation is responsible for the higher-energy part of the spectrum which
otherwise would not be felt by the s.p. potential. We may conclude that the
model algorithm rather carefully takes into account also the influence of the
collective dynamics at the higher angular momenta, which reflects on the
overall deformation characteristics of the CQOM potential.

Finally, special attention is given below to the behaviour of the Coriolis
mixing constant $A$ which, as already mentioned, plays an important role in
the formation of the IS energy and electromagnetic properties. The
calculations were repeated for four pairs of ($q_s,q_R$) values of the
quenching factors considered for the spin and collective gyromagnetic rates.
As the analysis of magnetic moments made in
Ref.~\cite{Minkov_Palffy_PRL_2019} suggests the need of rather strong
attenuation of the latter, here we consider $q_s$ and $q_R$ with slightly
lower values compared with the lowest pair ($q_s,q_R$)=($0.6,0.6$) considered
in Ref.~\cite{Minkov_Palffy_PRL_2019}. Thus, in the present calculations the
gyromagnetic quenching factors were allowed to be as small as
($q_s,q_R$)=($0.55,0.45$). Although the experimental value of the isomeric
energy is obviously out of reach for the model accuracy, we also consider its
theoretical prediction $E(3/2^{+})_{IS}$ in order to see how the model fits
``feel'' its tiny  energy scale as well as to assess accordingly the
relevance of the overall model description for the different deformations
within the model space.

\begin{figure*}[ht]
\centering
\includegraphics[width=8cm]{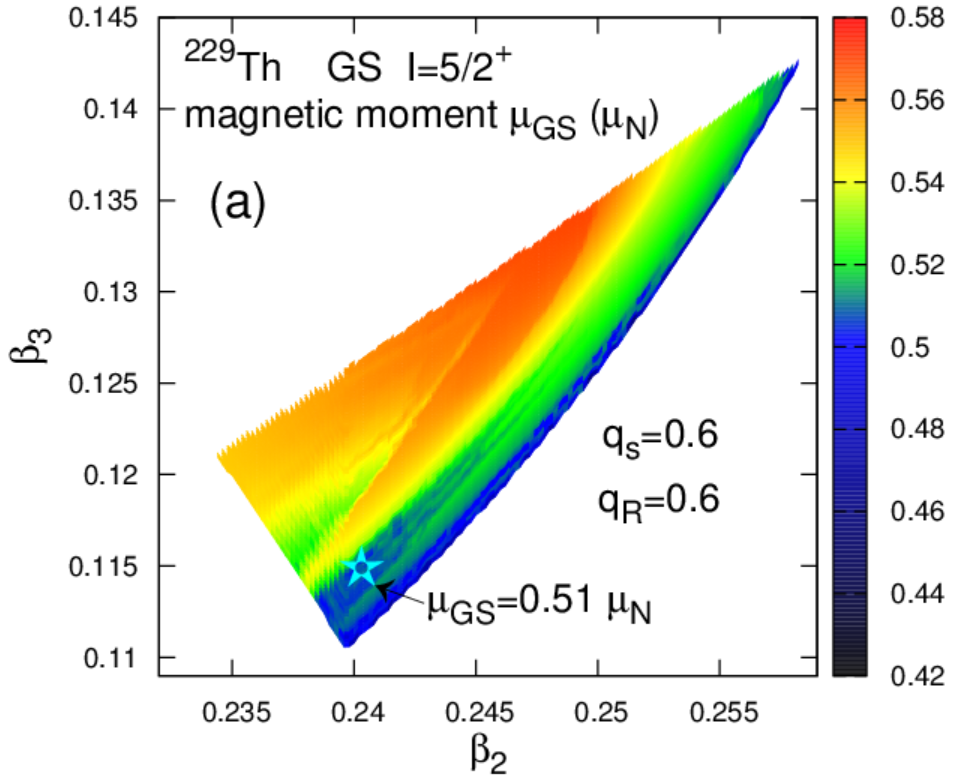}
\includegraphics[width=8cm]{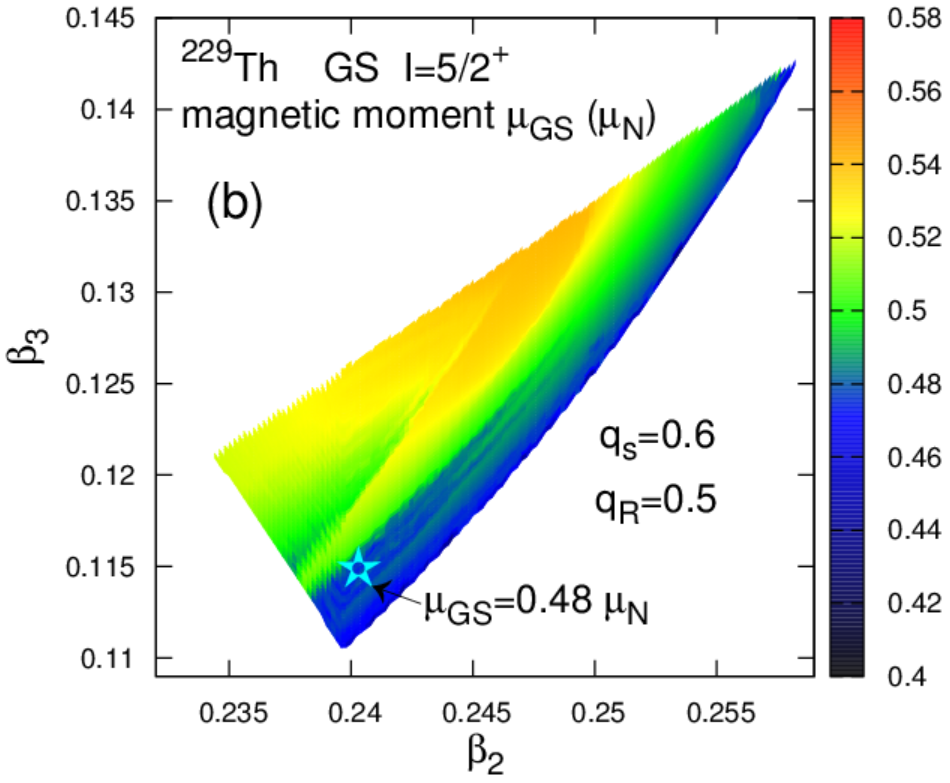}
\includegraphics[width=8cm]{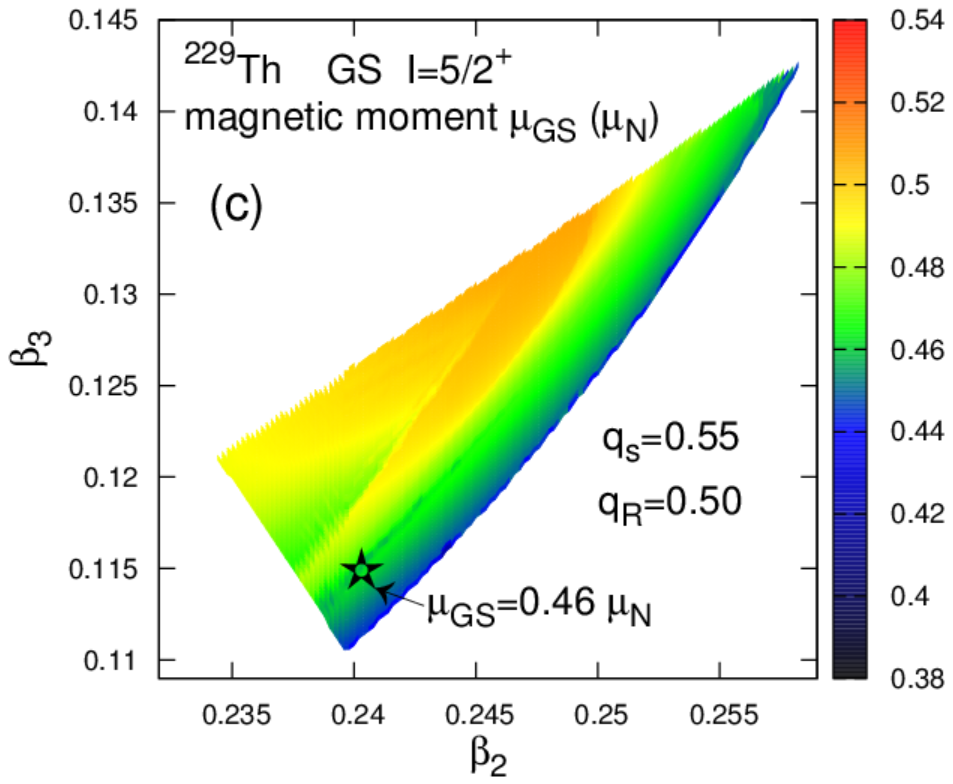}
\includegraphics[width=8cm]{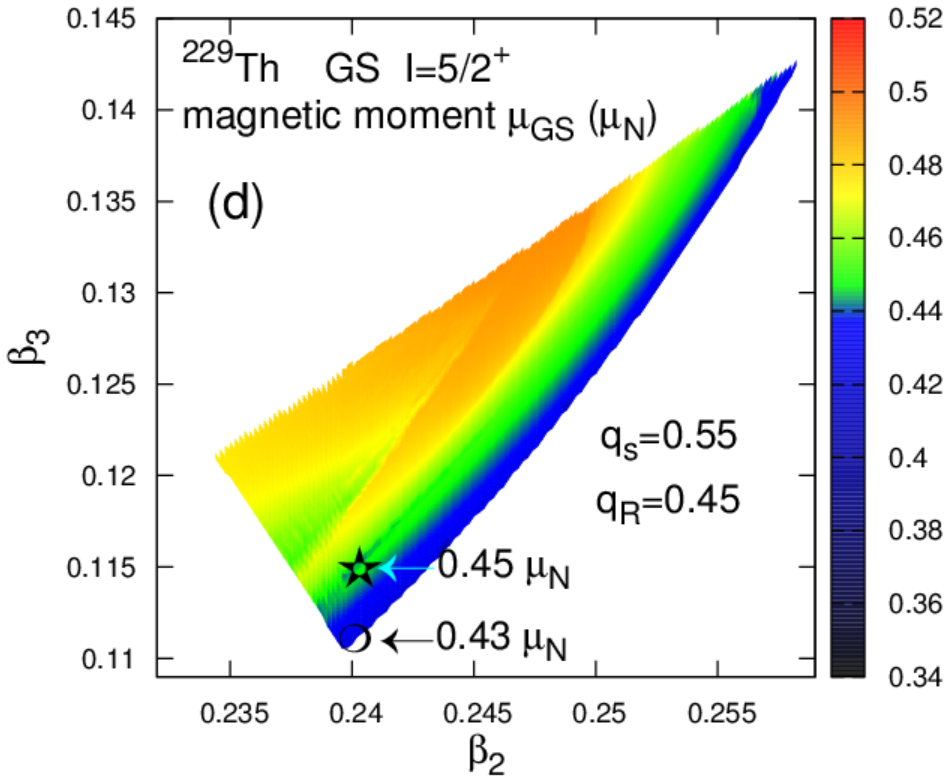}
\caption{GS magnetic moment values obtained by the model fits on a grid within
the model-defined QO deformation space at four different combinations of $q_s$
and $q_R$. The open star and the circle indicate the same sets of deformations
as shown in Figs.~\ref{th229_m1isotrans} and \ref{th229_e2isotrans}.}
\label{th229_gs_mdm}
\end{figure*}

\subsection{Energy description}

The primary quantity providing overall information about the relevance of the
model descriptions in the different points of the deformation space is the
root-mean-square (RMS) deviation between the theoretical and experimental
energy levels. Its behaviour as a function of the DSM QO deformation for
calculations made with quenching factors ($q_s,q_R$)=($0.6,0.6$) is
illustrated in Fig.~\ref{th229_rms}. We indicate with an open star the
location of the deformations ($\beta_2,\beta_3$)=($0.240,0.115$), adopted in
Refs.~\cite{Minkov_Palffy_PRL_2017,Minkov_Palffy_PRL_2019}. The RMS value
obtained in this point is about 34 keV which is the same as the value
obtained in the original model fits performed in
\cite{Minkov_Palffy_PRL_2017}, although now the experimental values of the
magnetic moments $\mu_{\mbox{\scriptsize GS}}$ and $\mu_{\mbox{\scriptsize
IS}}$ are included in the fits. We note that the RMS factor is close to this
value over a larger area of the deformation model space, demonstrating the
stability of the model solutions with the variation of QO deformations. The
upper part of the space with large $\beta_2$ and $\beta_3$ values, however,
appears unfavoured. We have verified that in all regions of the space with
the RMS close to 34 keV the description of the overall energy spectrum and
the available $B(M1)$ and $B(E2)$ transition rates is of similar accuracy as
the one reported in Ref.~\cite{Minkov_Palffy_PRL_2017}, with the obtained
CQOM parameter values being close to those in
Ref.~\cite{Minkov_Palffy_PRL_2017} (see Fig.~1 and Table~1 therein). We
notice that in the upper-left parts of the plot some lower RMS deviations are
obtained as low as $\approx$30 keV, however, for these deformations the model
predictions for the isomer energy are less favourable as analyzed below. In
addition, we found (barely visible in Fig.~\ref{th229_rms}) that towards the
line of the 5/2[633]--3/2[631] degeneracy the RMS factor sharply increases.
As discussed below in relation to the particular observables, this is the
result of the strong increase of the Coriolis $K$-mixing interaction which
largely exceeds the perturbation theory limitation and puts a constraint on
the model description valid close to the $5/2^{+}$--$3/2^{+}$ orbital
crossing.

In Fig.~\ref{th229_eniso} the theoretical isomer energy values $E_{IS}$
obtained by the model fits on the DSM QO space grid are presented for two
sets of gyromagnetic quenching values $(q_s,q_R)=(0.6,0.6)$ and
($0.55,0.45$). We find that for the first set the value obtained at the pair
of deformations ($\beta_2,\beta_3$)=($0.240,0.115$) is $E_{IS}\approx 1$ keV,
whereas for the second set it is $E_{IS}\approx 0.3$ keV. For the second set
we choose to demonstrate the result for one more pair of deformations from
our grid ($\beta_2,\beta_3$)=($0.2398,0.1108$), further on denoted for
simplicity by the rounded values ($0.240,0.111$), situated closer to the
degeneracy line. There we have $E_{IS}\approx 0.040$ keV already approaching
the scale of the experimental value. We note that for this pair of
deformations the $3/2^{+}$ q.p. energy yields $E^{\mbox{\scriptsize
qp}}_{3/2^{+}}=0.188$ keV.

All $E_{IS}$ values shown in Fig.~\ref{th229_eniso} are obtained in the
fitting procedure on the same footing without particular refinement. As
already mentioned, one can easily achieve the exact experimental value of
0.008 keV through a very fine tuning of model parameters (e.g. the $K$-mixing
$A$), with a minimal deterioration of the description in the remaining energy
levels (see also Ref.~\cite{Minkov_Palffy_PRL_2017}). The plots in
Fig.~\ref{th229_eniso} show that in the large areas of the model space the
fits provide reasonable values of $E_{IS}$ which could be renormalized to the
experiment in this manner. However, we also see that in the upper-left parts
of the plots the $E_{IS}$ considerably increases up to 7-8 keV giving an
indication that at these deformations the remoteness of the 5/2[633] and
3/2[631] orbitals (see Fig.~\ref{th229_spqpen}) already makes it difficult
for the model mechanism to achieve quasidegeneracy. Also, we notice a thin
stripe with large $E_{IS}$ values along the line of degeneracy which
obviously indicates the limitation of the perturbation theory. Concluding
this part, our analysis of the RMS factor and isomer energy values outlines
certain limits of reliability of the present model application and favours
the lower vertex of the model space around the deformation set used in
Ref.~\cite{Minkov_Palffy_PRL_2017} in reasonable proximity to the
$5/2^{+}$--$3/2^{+}$ orbitals crossing line.

\subsection{$B(M1)$ and $B(E2)$ isomer transition rates}

\begin{figure}[ht]
\centering
\includegraphics[width=8cm]{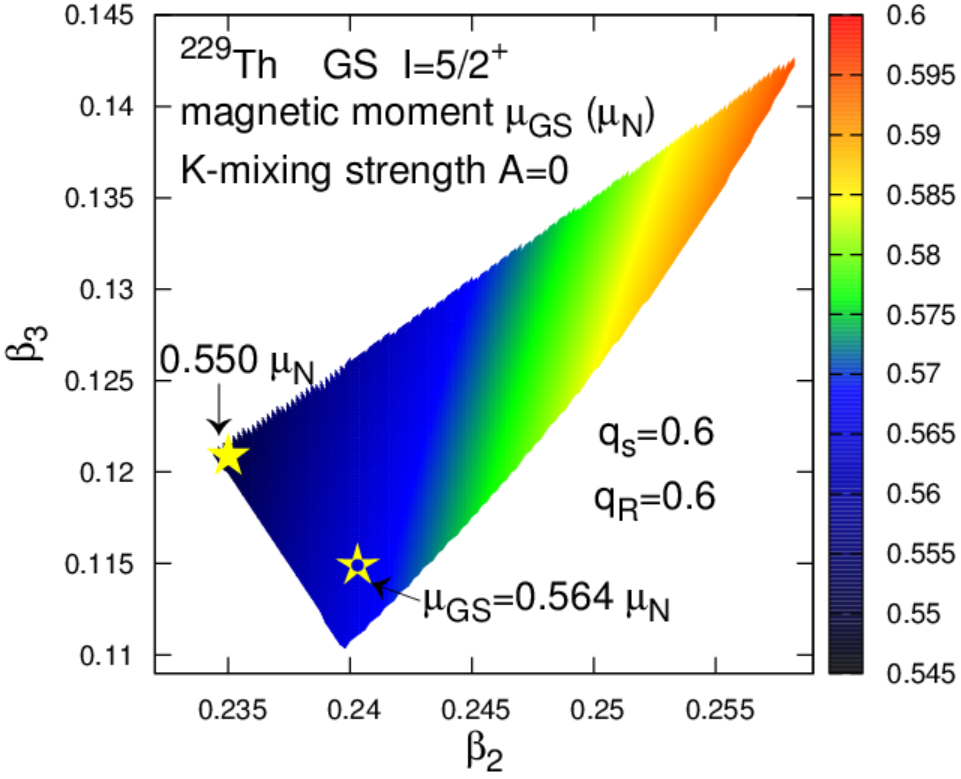}
\caption{The ``bare'' s.p. GS magnetic moment values obtained without Coriolis
mixing ($A=0$) on a grid within the model-defined QO deformation space with
$q_s=q_R=0.6$. The open star indicates the location of the deformations
($\beta_2,\beta_3$)=($0.240,0.115$) adopted in
Refs.~\cite{Minkov_Palffy_PRL_2017,Minkov_Palffy_PRL_2019}. The full star
indicates the grid point which provides the lowest value
$\mu_{\mbox{\scriptsize GS}}$=0.55 $\mu_{N}$.}
\label{th229_gs_mdm_bare}
\end{figure}

\begin{figure*}[ht]
\centering
\includegraphics[width=8cm]{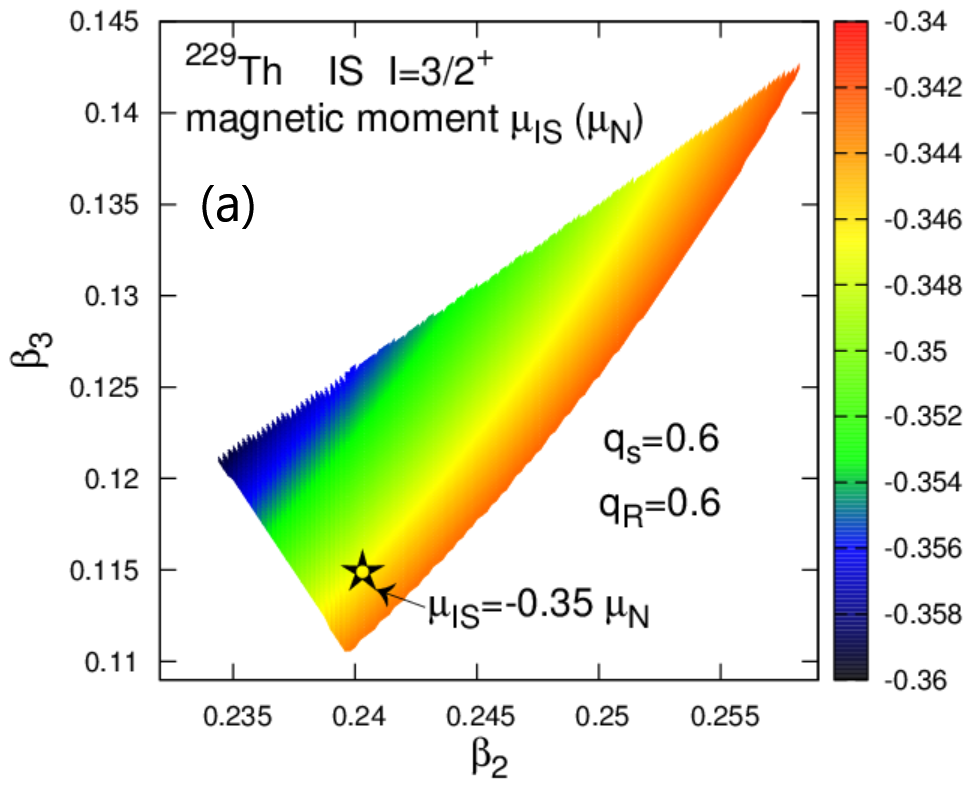}
\includegraphics[width=8cm]{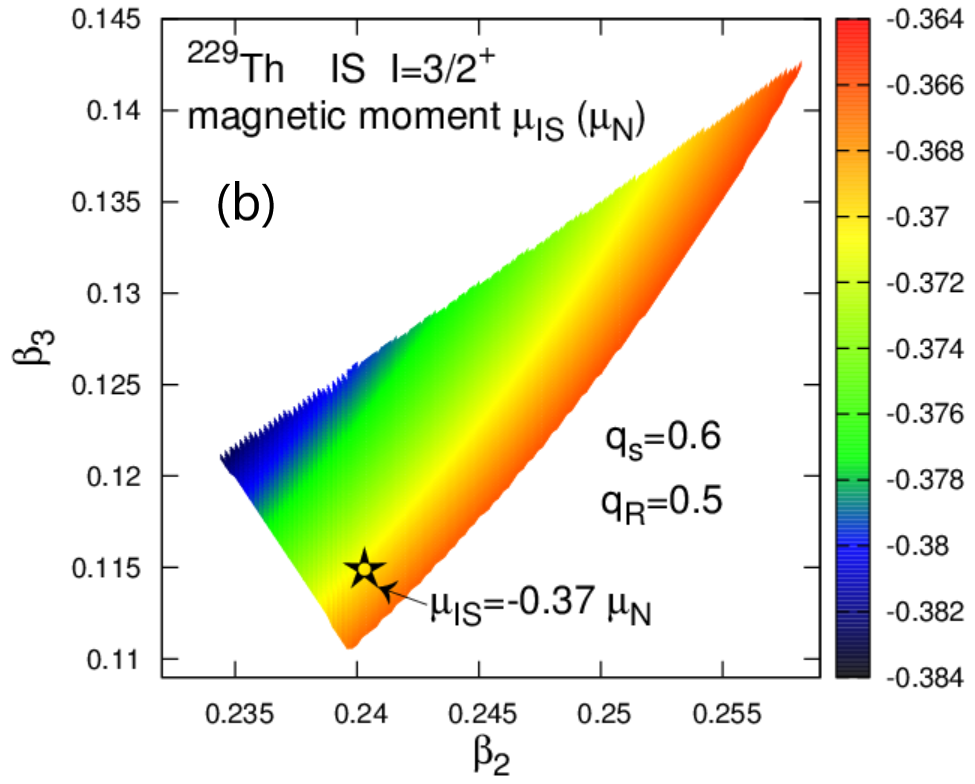}
\includegraphics[width=8cm]{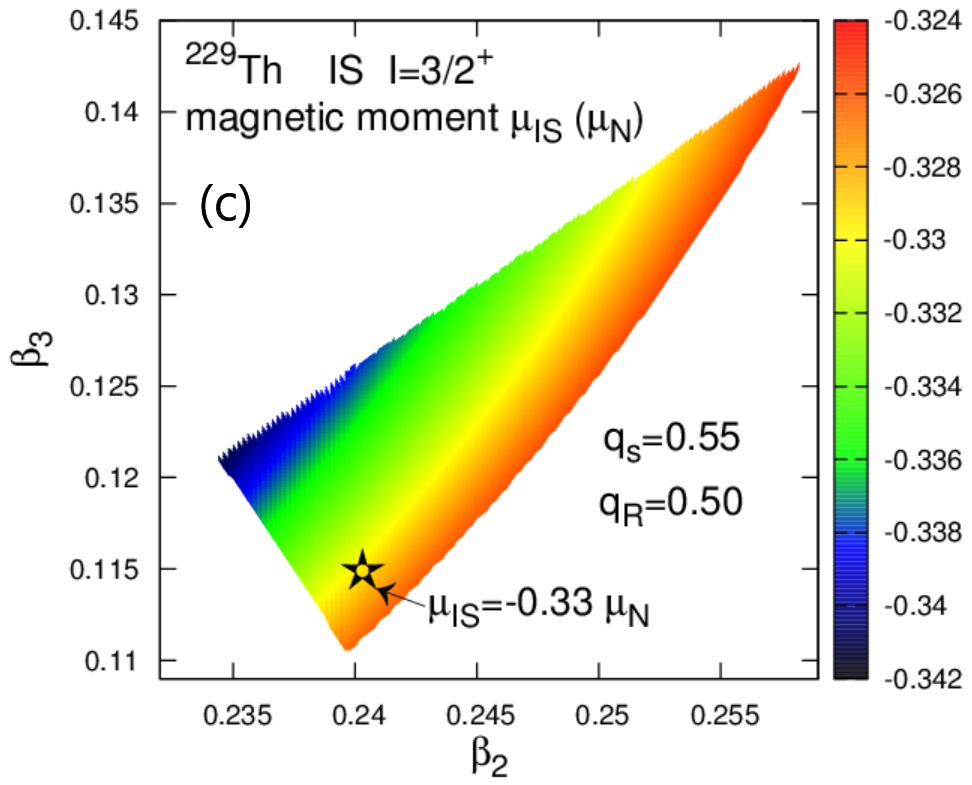}
\includegraphics[width=8cm]{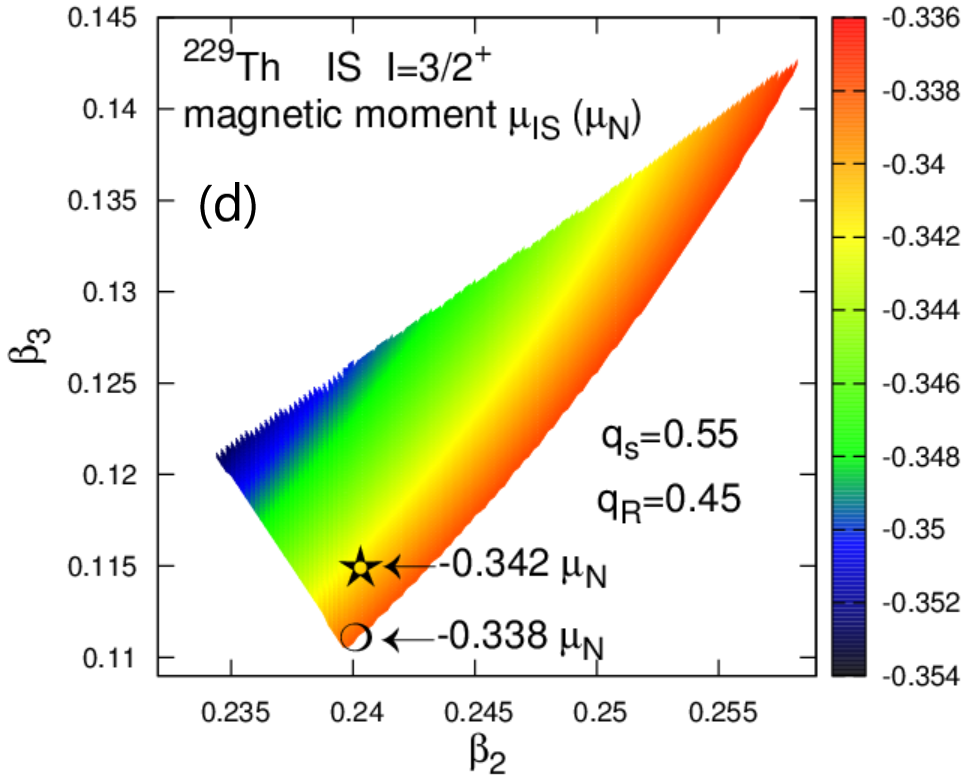}
\caption{IS magnetic moment values obtained by the model fits on the
model-defined QO deformation space grid at four different combinations of
$q_s$ and $q_R$. The open star and the circle indicate the same sets of
deformations as shown in Fig.~\ref{th229_gs_mdm}.}
\label{th229_is_mdm}
\end{figure*}

The results obtained for the isomeric $B(M1;\, 3/2^{+}_{\mbox{\scriptsize
IS}}\rightarrow 5/2^{+}_{\mbox{\scriptsize GS}}$) and $B(E2;\,
3/2^{+}_{\mbox{\scriptsize IS}}\rightarrow 5/2^{+}_{\mbox{\scriptsize GS}}$)
transition rates are illustrated in Figs.~\ref{th229_m1isotrans} and
\ref{th229_e2isotrans}, respectively, for the four sets of quenching factors.
As in the energy analysis above, here we mark with an open star the value
obtained by the model at the pair of QO deformations
($\beta_2,\beta_3$)=($0.240,0.115$), adopted in
Refs.~\cite{Minkov_Palffy_PRL_2017,Minkov_Palffy_PRL_2019}, for which the
original model predictions for the transition rates and magnetic moments were
made. Also, we examine the model predictions towards the line of
$5/2^{+}$--$3/2^{+}$ orbitals degeneracy by considering in the graphs of
($q_s,q_R$)=($0.55,0.45$) the pair of deformations
($\beta_2,\beta_3$)=($0.240,0.111$) indicated by the open circle. Inspecting
Fig.~\ref{th229_m1isotrans}, first we observe that the overall behaviour of
the $B(M1;\, 3/2^{+}_{\mbox{\scriptsize IS}}\rightarrow
5/2^{+}_{\mbox{\scriptsize GS}}$) transition value shows an increase with the
approaching of the degeneracy line. This is due to the circumstance that with
the decreasing distance between both orbitals $5/2^{+}$ and $3/2^{+}$, the
$K$-mixing effect generated by the matrix element in Eq.~(\ref{amix}) sharply
increases and this leads to the increase in the connecting transition rates.
It should be noted, however, that in the model procedure this increase is
counterbalanced by the adjustable parameter $A$ which drops accordingly thus
preventing a deterioration of the model description due to the excessive
mixing force. This will be discussed in more detail in the following (see
Fig.~\ref{th229_amix} and related text below). Keeping in mind this
clarification, we notice in Fig.~\ref{th229_m1isotrans}(d) that while the
quenching of the gyromagnetic factors leads to a reduction of
$B(M1)_{\mbox{\scriptsize IS}}$ to 0.005 W.u. at
($\beta_2,\beta_3$)=($0.240,0.115$) (an effect already addressed in
Ref.~\cite{Minkov_Palffy_PRL_2019}), the shift of the deformation towards the
degeneration line returns the value back to 0.007 W.u., i.e. in the original
range of the prediction made in Ref.~\cite{Minkov_Palffy_PRL_2017}.

A similar behaviour of the $B(E2;\, 3/2^{+}_{\mbox{\scriptsize
IS}}\rightarrow 5/2^{+}_{\mbox{\scriptsize GS}}$) transition rate in the
model deformation space is observed in Fig.~\ref{th229_e2isotrans}. We note
that here all obtained $B(E2)_{\mbox{\scriptsize IS}}$ values exceed the
previous prediction \cite{Minkov_Palffy_PRL_2017} but stay in the range of
the correction suggested in Ref.~\cite{Minkov_Palffy_PRL_2019}.

\subsection{Ground-state and isomer magnetic moments}

The calculated GS  magnetic moment $\mu_{\mbox{\scriptsize GS}}$ for the four
sets of quenching factors ($q_s,q_R$) is illustrated in
Fig.~\ref{th229_gs_mdm}. The overall model behaviour of this quantity is such
that it decreases both with the attenuation of the gyromagnetic factors and
with the approaching of the $5/2^{+}$--$3/2^{+}$ orbital-degeneracy line. We
see that for ($\beta_2,\beta_3$)=($0.240,0.115$) it drops to 0.45 $\mu_{N}$,
while with the shift of the deformations to the point ($0.240,0.111$) it
reaches the value of 0.43 $\mu_{N}$. The latter result is obviously due to
the increasing Coriolis mixing which reduces the value of
$\mu_{\mbox{\scriptsize GS}}$ as has been shown already in
Ref.~\cite{Minkov_Palffy_PRL_2019}. However, it appears that for model
conditions considered physically reasonable, this is still not sufficient to
reproduce the newer experimental value of 0.360(7) $\mu_{N}$
\cite{Safronova13}, although the model reproduces fairly well the earlier
measured value of 0.46(4) $\mu_{N}$ \cite{Gerstenkorn74}.

\begin{figure*}[ht]
\centering
\includegraphics[width=8cm]{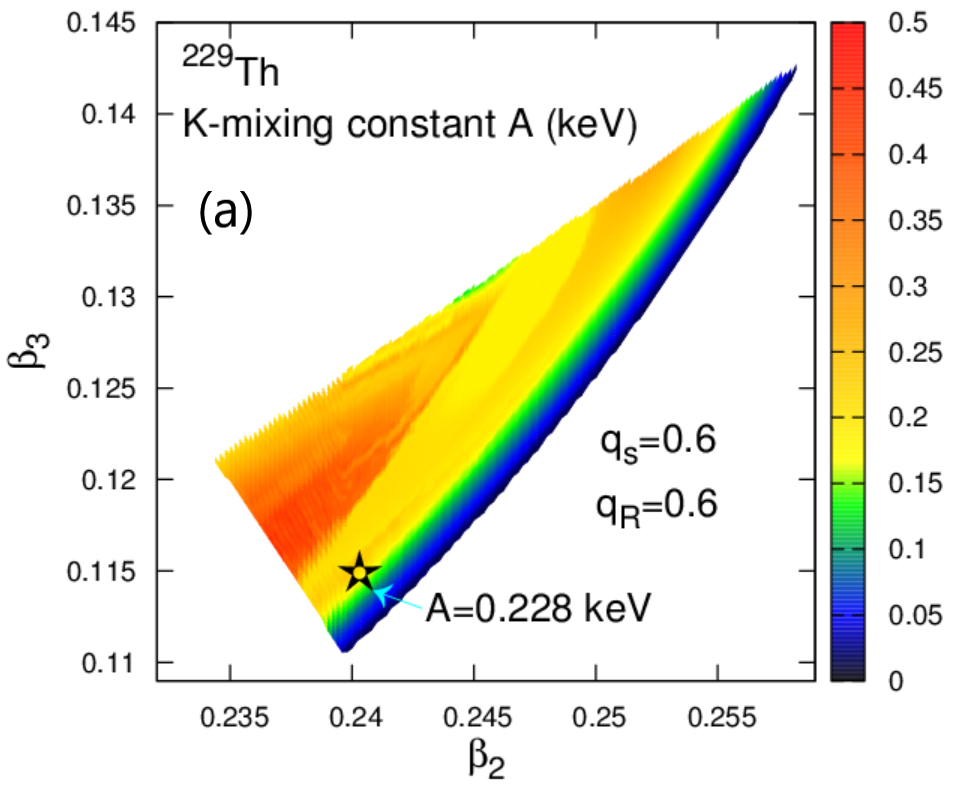}
\includegraphics[width=8cm]{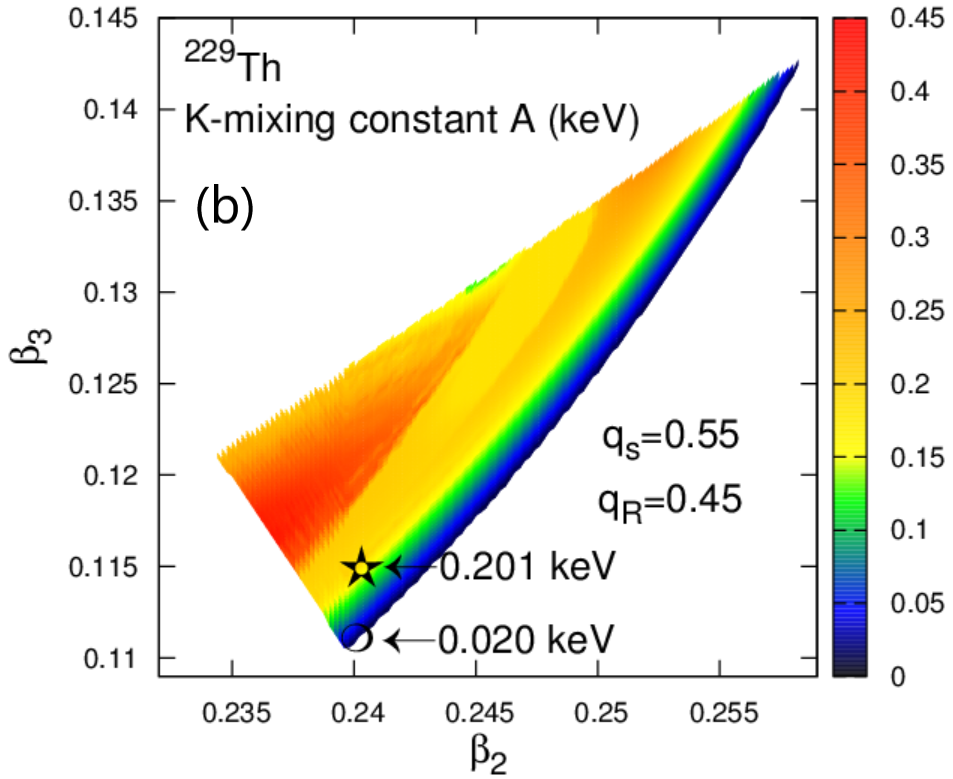}
\caption{Values of the $K$-mixing constant $A$ obtained by the model fits on the
model-defined QO deformation space grid at two different combinations of $q_s$
and $q_R$. The open star and the circle indicate the same sets of deformations
as shown in Fig.~\ref{th229_gs_mdm}.}
\label{th229_amix}
\end{figure*}

It is instructive to check here also the ``bare'' values of
$\mu_{\mbox{\scriptsize GS}}$, i.e. those obtained by the pure s.p. wave
function without including the Coriolis mixing. Therefore, in
Fig.~\ref{th229_gs_mdm_bare} we show the analog of Fig.~\ref{th229_gs_mdm}(a)
(with $q_s=q_R=0.6$) in which $\mu_{\mbox{\scriptsize GS}}$ is calculated in
the absence of Coriolis mixing with the $K$-mixing constant $A=0$. In this
case Eq.~(\ref{mumod}) reduces to the terms in its first line with the second
term involving the expression of Eq.~(\ref{gKb}). Here we first see that
$\mu_{\mbox{\scriptsize GS}}$ appears with considerably larger values in the
limits $\mu_{\mbox{\scriptsize GS}}$=0.55--0.60 $\mu_{N}$ which also show
different behaviour in the DSM QO space compared with the Coriolis-mixing
case. This result does not depend on the model-parameters fit and illustrates
the genuine contribution of the QO deformation for the formation of the GS
magnetic moment of $^{229}$Th. Comparing both plots we see that in the pure
s.p. case without Coriolis mixing, the lowest $\mu_{\mbox{\scriptsize
GS}}=0.55$ $\mu_{N}$ value appears in the left-upper vertex of the triangle
model space, whereas in the Coriolis-mixing case the low values (lower than
the pure s.p. ones), appear in the lower vertex of the space.

The above result leads us to the following conclusions. The Coriolis effect
causes a decrease of the nuclear magnetic moment in the $^{229}$Th GS
throughout the model deformation space. It plays a considerable role in our
approach for fixing the GS magnetic moment through the overall model fits,
although this is still not enough to reproduce the latest adopted
experimental value.  The appearance of essentially lower
$\mu_{\mbox{\scriptsize GS}}$-values obtained through the adjusted $K$-mixing
constant $A$ compared with the corresponding pure s.p.
$\mu_{\mbox{\scriptsize GS}}$-values shows that the increase of the
model-controlled Coriolis-mixing towards the $5/2^{+}$--$3/2^{+}$
orbitals-degeneracy line essentially determines the behaviour of the GS
magnetic moment and dominates over the corresponding effect of changing QO
deformation on the pure s.p. $\mu_{\mbox{\scriptsize GS}}$-values. This
conclusion suggests that no considerably different result can be reached
through further variation of deformations parameters in the DSM QO space.

Figure~\ref{th229_is_mdm} shows the calculated IS magnetic moment
$\mu_{\mbox{\scriptsize IS}}$ for the four sets of gyromagnetic quenching
factors considered ($q_s,q_R$). We note its relatively flat behaviour as a
function of the QO deformation, with a slight increase towards the degeneracy
line. Since $\mu_{\mbox{\scriptsize IS}}$ is practically not affected by the
$K$-mixing effect, we may claim that this dependence can be considered as the
bare effect of the changing structure of the s.p. wave functions along the
deformation space. We see that in all plots of Fig.~\ref{th229_is_mdm} the
lowest value of $\mu_{\mbox{\scriptsize IS}}$ appears in the left upper
corner of the model space similarly to the ``bare'' (s.p.)
$\mu_{\mbox{\scriptsize GS}}$ case (Fig.~\ref{th229_gs_mdm_bare}). For
example in the case of $q_s=q_R=0.6$, Fig.~\ref{th229_is_mdm}(a), the
corresponding lowest value $\mu_{\mbox{\scriptsize IS}}$=$-$0.36 $\mu_{N}$
comes closer to the experimental result. However, the model fits have shown
that the lower corner provides better predictions for $\mu_{\mbox{\scriptsize
GS}}$, and we keep our attention on this region. Besides, for all considered
quenching factors ($q_s , q_R$),  the theoretical $\mu_{\mbox{\scriptsize
IS}}$ values appearing in Fig.~\ref{th229_is_mdm} enter the error bars of the
recent experimental value of $-$0.37(6) $\mu_{N}$ reported in
Ref.~\cite{Thielking2018,Mueller18}.

\begin{figure*}[ht]
\centering
\includegraphics[width=8cm]{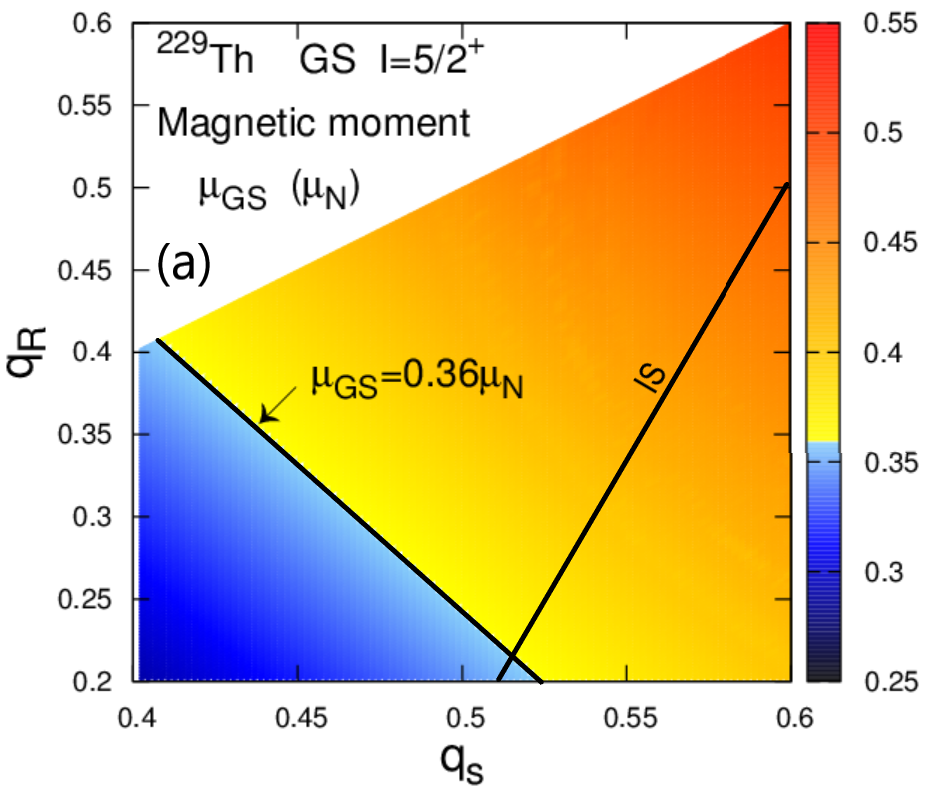}
\includegraphics[width=8cm]{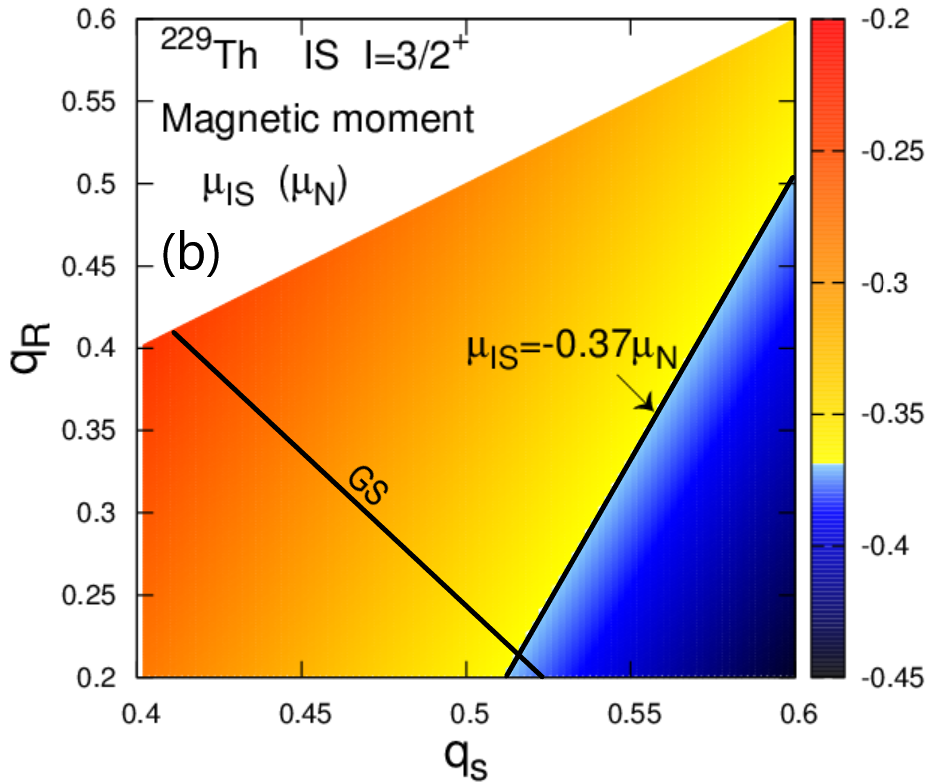}
\caption{(a) GS and (b) IS magnetic moment values obtained by the model fits
on a grid for the spin-gyromagnetic, $q_s$, and rotation-gyromagnetic, $q_R$,
attenuation factors. The black lines show the values that would reproduce
the corresponding GS and IS experimental values.}
\label{th229_mdm_qsqr}
\end{figure*}

The results presented so far already reveal important details and relations
characterizing the model mechanism upon which the $^{229m}$Th isomer is
formed and its spectroscopic properties develop. Obviously the proximity of
the 5/2[633] and 3/2[631] s.p. orbitals plays a major role providing the
overall condition for the appearance of a low-lying excitation. Now this is
clearly quantified by all above plots. On the other hand, it is also clear
that the appearance of the isomer can not be due only to the orbital
quasidegeneracy. The reason is that at the distance of few eV the mixing
between the two orbitals becomes very large and pushes all related
observables in unphysical regions of magnitude. In this case the perturbation
terms in the centrifugal part, Eq.~(\ref{Xmix}), of the Hamiltonian as well
as in the Coriolis perturbed wave function, Eq.~(\ref{wtcoriol}), collapse in
a singularity. The model prevents this situation mostly through the $K^2$
term and $K$-mixing constant $A$ in Eq.~(\ref{Xmix}), which allow us to
properly situate both GS and IS, i.e., to obtain the IS energy value as small
as necessary, by keeping the 3/2[631]--5/2[633] orbital distance aside from
the degeneracy line. In this respect one can say that the physically adequate
QO deformations are slightly aside from this line.

The model mechanism feature described above can be seen by following the
behaviour of the Coriolis mixing constant $A$ adjusted at each grid point in
the deformation space. We investigate this in Fig.~\ref{th229_amix} for two
sets of quenching factors, $(q_s,q_R)=(0.6,0.6)$ and $(0.55,0.45)$.  The
obtained values of $A$ range from zero to approximately 0.5 keV. Towards the
degeneracy line, where the $K$-mixing matrix element in Eq.~(\ref{amix})
connecting the two orbitals sharply increases as they approach each other,
the adjustment algorithm strongly reduces the value of $A$. In this way the
model ``feels'' the growing magnitude of the Coriolis mixing and tries to
compensate its excessive effect on the considered observables through the
parameter $A$ keeping them in physically meaningful ranges. Providing this
balancing role of the parameter $A$ and having in mind all so far obtained
model patterns for the spectroscopic observables in $^{229}$Th we can be
rather confident in the consistency of the analysis made and the reliability
of the QO deformation region outlined.

Finally, it is interesting to identify the degree of spin and collective
gyromagnetic factor attenuations required to reproduce in the present model
both experimental $\mu_{\mbox{\scriptsize GS}}$ and $\mu_{\mbox{\scriptsize
IS}}$ values. This is shown in Fig.~\ref{th229_mdm_qsqr}, where the values of
each of these quantities obtained in the model fits at
($\beta_2,\beta_3$)=($0.240,0.115$) are given [Fig.~\ref{th229_mdm_qsqr}(a)
for $\mu_{\mbox{\scriptsize GS}}$ and Fig.~\ref{th229_mdm_qsqr}(b) for
$\mu_{\mbox{\scriptsize IS}}$] as functions of the quenching factors $q_s$
and $q_R$. The black lines denote the pairs of ($q_s,q_R$) values which
provide the corresponding $\mu_{\mbox{\scriptsize GS}}$ and
$\mu_{\mbox{\scriptsize IS}}$ experimental values. The crossing of both lines
shows the point at which both magnetic moments are reproduced together. We
see that this occurs at $q_s\approx 0.52$ and a rather low value of
$q_R\approx 0.22$, which corresponds to a quite strong attenuation of the
collective gyromagnetic factor. Because this quenching magnitude
 is hard to justify, we conclude that the agreement
between the present theoretical model and the currently adopted experimental
value of the GS magnetic moment remains an open issue.

\section{Summary and conclusion}
\label{summ}

In this work we have thoroughly examined the physical conditions for the
formation of the 8 eV isomer of $^{229}$Th according to the model mechanism
suggested by our QO vibration-rotation core plus particle approach. First, we
have determined the model deformation space encompassing the Woods-Saxon DSM
quadrupole and octupole deformations which allow the appearance of the GS and
IS with the experimentally adopted $K$-values and parities. We were able to
clearly identify its borders constrained by the average parity of the isomer
and the crossings of the 3/2[631] orbital on which the IS is built with the
5/2[633] orbital of the GS as well as with a 7/2[743] orbital. This space is
rather limited and essentially constrains the QO deformation in the s.p.
potential within the ranges $0.235\leq\beta_{2}\leq 0.255$ and
$0.11\leq\beta_{3}\leq 0.14$. These results lead us to the important
conclusion that the appearance of the $K^{\pi}=3/2^{+}$ IS through this
mechanism is only possible in the presence of essentially nonzero octupole
deformation in the s.p. potential.

Furthermore, we have examined the dependence of the overall DSM+CQOM model
description on the QO deformations within the DSM model space. Our analysis
of the CQOM fits made in Sec.~\ref{cqomfits}, including the CQOM
potential-bottom semiaxes and dynamical deformations, showed that the
collective CQOM conditions under which the $^{229m}$Th isomer is formed
consistently interrelate with the relevant conditions provided by the
odd-nucleon degrees of freedom. Under these overall conditions we obtain for
all of the considered observables, $B(M1;\, 3/2^{+}_{\mbox{\scriptsize
IS}}\rightarrow 5/2^{+}_{\mbox{\scriptsize GS}}$), $B(E2;\,
3/2^{+}_{\mbox{\scriptsize IS}}\rightarrow 5/2^{+}_{\mbox{\scriptsize GS}}$),
$\mu_{\mbox{\scriptsize GS}}$, $\mu_{\mbox{\scriptsize IS}}$ and
$E(3/2^{+})_{IS}$ a smooth behaviour of the model predictions and
descriptions compared with the results obtained for the fixed pair of
Woods-Saxon DSM QO deformations considered in our previous works
\cite{Minkov_Palffy_PRL_2017, Minkov_Palffy_PRL_2019}, with the only
peculiarity appearing close to the line of $5/2^{+}$--$3/2^{+}$ degeneracy
where the mixing of both orbitals exceeds the perturbation theory limits. On
the other hand the corresponding behaviour of the model energy RMS factors
and Coriolis mixing constant shows that descriptions obtained with values of
the above observables essentially deviating from those obtained in
\cite{Minkov_Palffy_PRL_2017, Minkov_Palffy_PRL_2019} are of lower quality
and/or violate the perturbation limit. For the remaining descriptions we have
verified that in the limits of moderate deviations of the considered
observables from the original values in \cite{Minkov_Palffy_PRL_2017,
Minkov_Palffy_PRL_2019}, the model is renormalizable, so that through a small
variation of model parameters and on the expense of small deteriorations of
the RMS factor, we can get very similar model predictions for the different
pairs of QO deformations. Using this model feature as well as assuming
possible stronger attenuation of the spin and collective gyromagnetic
factors, we have outlined rather narrow limits of arbitrariness in which the
model values of each of the above quantities can vary by keeping its
reasonable physical meaning and predictability.

Our main conclusion is that within the obtained model deformation space the
applied QO core plus particle approach provides a rather constrained
prediction for the most important $^{229}$Th energy and electromagnetic
characteristics related to the formation and manifestation of the 8 eV
isomer. This allows us to generally reconfirm the predictions initially made
in Ref.~\cite{Minkov_Palffy_PRL_2017, Minkov_Palffy_PRL_2019} and to
summarize them with a slight update: The $B(M1)$ IS transition remains in the
limits 0.006-0.008 W.u. with an open possibility towards lower values such as
0.005 W.u.; the $B(E2)$ IS transition may be considered with slightly higher
values between 30 and 50 W.u., compared with those in
Ref.~\cite{Minkov_Palffy_PRL_2017}; the GS magnetic moment allows a limited
possibility for variation and remains with a model value around 0.50
$\mu_{N}$ possibly getting values as smaller as 0.43-0.48 $\mu_{N}$ under a
stronger assumption for the gyromagnetic factors attenuation, thus covering
the old experimental value of Ref~\cite{Gerstenkorn74}, but still
overestimating the newer one of Ref.~\cite{Safronova13}; the theoretical IS
magnetic moment firmly reproduces the recent experimental value within the
uncertainty limits reported in \cite{Thielking2018,Mueller18} and this is
obtained under all considered model conditions; and finally the model values
for the isomer energy typically obtained around 1 keV and below suggest that
with a small variation of parameters, and with the expense of a minor
deterioration of the other energy levels, the model can easily reproduce the
experimental value, although this is of little importance due to the overall
limitation of the model accuracy in the energy-spectrum description.

We note that for all of the above quoted values the other model observables
(energy levels and transition rates), for which experimental data are
available, remain described within the accuracy limits reported in
Ref.~\cite{Minkov_Palffy_PRL_2017}. Thus our analysis suggests that within
the above outlined limits of arbitrariness the model predictions could
provide reliable estimates for the $^{229m}$Th spectroscopic characteristics
which could serve to the experiment in further efforts to observe and control
the yet elusive radiative isomer transition.

Finally, the results obtained confirm the relevance of the model mechanism
emphasizing on the role of the fine interplay between nuclear collective and
intrinsic degrees of freedom as a plausible reason for the isomer formation.
On this basis we conclude that the same dynamical mechanism may govern also
in other nuclei the formation of excitations close to the border of atomic
physics energy scale. Such states may exist being not yet observed due to
experimental difficulties similar to those encountered in $^{229m}$Th. As in
this work we give a detailed prescription about the examination and
constraining of the physical conditions under which such a phenomenon may
emerge, it appears promising to extend the study to other nuclei in the same
or other mass regions. In this aspect the close neighbour $^{231}$Th as well
as the $^{235m}$U isomer would be natural candidates for such a study. This
could be a subject of future work.

\section*{ACKNOWLEDGMENTS}

This work is supported by the Bulgarian National Science Fund (BNSF) under
Contract No. KP-06-N48/1. AP gratefully acknowledges funding by the EU
FET-Open project 664732 (nuClock). This work is part of the
ThoriumNuclearClock project that has received funding from the European
Research Council (ERC) under the European Union’s Horizon 2020 research and
innovation programme (Grant agreement No. 856415).


\bibliography{refs}

\end{document}